\documentclass{iopart}
\usepackage{iopams}
\usepackage{graphics}
\usepackage{graphicx}
\usepackage{color}
\usepackage{braket}
\usepackage{bm}
\usepackage{float}
\usepackage{hyperref}
\usepackage[numbers,sort&compress]{natbib}
\usepackage{dsfont}
\newenvironment{myproof}{\par\noindent\textbf{Proof.}\ }{\hfill$\square$\par}

\newcommand{\be}{\begin{equation}}\newcommand{\ee}{\end{equation}}
\newcommand{\bea}{\begin{eqnarray}}\newcommand{\eea}{\end{eqnarray}}
\newcommand{\brr}{\begin{array}}\newcommand{\err}{\end{array}}
\newcommand{\bit}{\begin{itemize}}\newcommand{\eit}{\end{itemize}}
\newcommand{\ben}{\begin{enumerate}}\newcommand{\een}{\end{enumerate}}

\newcommand{\ba}{\begin{array}}
\newcommand{\ea}{\end{array}}

\def\1{{_{1}}}\def\2{{_{2}}}

\newcommand{\eqref}[1]{(\ref{#1})}

\begin{document}
\title[Geometric measures of quantum nonlocality]{Geometric measures of quantum nonlocality: characterization, quantification, and comparison by distances and operations}

\author{
Gennaro Zanfardino$^{1,2,3}$, 
Wojciech Roga$^4$, 
Gianluigi Tartaglione$^{1}$,
Masahiro Takeoka$^4$,
and Fabrizio Illuminati$^{2,1}$
}

\address{$^1$ Dipartimento di Ingegneria Industriale, Universit\`a degli Studi di Salerno, Via Giovanni Paolo II, 132 I-84084 Fisciano (SA), Italy}
\address{$^2$ Institute of Nanotechnology of the National Research Council of Italy, CNR-NANOTEC, Lecce Unit, Via Monteroni 165, I-00173, Lecce, Italy}
\address{$^3$ INFN, Sezione di Napoli, Gruppo collegato di Salerno, Italy}
\address{$^4$ Department of Electronics and Electrical Engineering, Graduate School of Science and Technology, Keio University, Yagami Campus, 3-14-1 Hiyoshi, Kohoku-ku, Yokohama, Kanagawa 223-8522, Japan}

\eads{\mailto{gzanfardino@unisa.it}, \mailto{wojciech.roga@keio.jp}, \mailto{gtartaglione@unisa.it}, \mailto{takeoka@elec.keio.ac.jp}, 
\mailto{filluminati@unisa.it}}

\date{\today}

\begin{abstract}
We introduce a geometric framework for studying Bell nonlocality in Hilbert space, where, for a given quantum state, nonlocality is quantified by the distance between the state and the set of local states. This approach applies to any Bell inequality and any measurement scenario. Whenever the local set is characterized, the proposed nonlocality measure can be computed explicitly.

As a general result, we prove that for any scenario in arbitrary dimension the closest local state to a Werner state is itself a Werner state, and analogously, the closest local state to an isotropic state is again isotropic. In the two-qubit case, we further show that the closest local state to a Bell-diagonal state is Bell-diagonal as well. These structural results are independent of the specific Bell inequality considered, thus revealing intrinsic geometric features of these families of states and providing significant simplifications for computing the proposed measures.

For the Clauser–Horne–Shimony–Holt (CHSH) inequality in two-qubit systems and the Collins–Gisin–Linden–Massar–Popescu (CGLMP) inequality for two qudits of arbitrary finite dimension, we derive explicit geometric measures of nonlocality for Bell-diagonal, Werner, and isotropic states using various distance metrics, including the trace, Hellinger, Hilbert–Schmidt distances, and relative entropy. Furthermore, we prove in all generality that for all scenarios in which the local set is not fully characterized, the geometric measures provide rigorous lower bounds on nonlocality.

\end{abstract}

\maketitle

\vskip -1.0 truecm 

\maketitle

\section{Introduction}

Quantum nonlocality is generally understood as the impossibility of describing experimental correlations using a local hidden variable (LHV) model \cite{Bell1964, Brunner2014}. A multipartite state is considered nonlocal if there exists at least one experimental configuration, involving non-communicating parties that can at most share randomness, which leads to a violation of a Bell-type inequality.

It is clear that a state that appears local with respect to one Bell inequality may still violate a different Bell inequality \cite{Collins2004,Laskowski2004, Masanes2008}. It can also happen that a local quantum state can violate a Bell inequality after being preprocessed by local filtering operations, revealing a \textit{hidden nonlocality} \cite{Popescu1995, Gisin1996}.  Furthermore, there exist quantum states that do not violate any Bell inequality individually, but when multiple copies of the state are considered along with joint measurements, the resulting system can exhibit a violation of some Bell inequality. This phenomenon is known as \textit{superactivation} of Bell nonlocality \cite{Navascues2011,Palazuelos2012}.  

Determining whether a state is local, that is, whether its correlations admit a local hidden variable (LHV) description, can be challenging as it requires testing the state against an enormous number of possible experimental scenarios. For example, it has been shown that there exist two-qubit Werner states \cite{Werner1989} that do not violate the Clauser–Horne–Shimony–Holt (CHSH) inequality \cite{Clauser1969} but nevertheless exhibit nonlocality when considering more than two measurements per party \cite{Vertesi2008}. Therefore, for a fixed scenario involving $n$ observers, each allowed to perform $m_{i}$ measurements labeled by $x_{i}$ ($i = 1, \dots, n$), and each measurement having $\Delta_{i}$ possible outcomes $a_{i}$, we define the set $\mathcal{L}_{\vec{m}, \vec{\Delta}}$ of \textit{local states}, i.e. the set of states that admit a local hidden variable (LHV) description, as follows:
\begin{equation}
    p(\vec{a}|\vec{x}) = \sum_{\lambda} p(\lambda) \prod_{i=1}^{n} p(a_{i}|\lambda, x_{i}),
    \label{local_description}
\end{equation}
for any choice of $\vec{a}$ and $\vec{x}$, where
\begin{equation}
    p(\vec{a}|\vec{x}) = {\rm Tr}\!\left[\rho \bigotimes_{i=1}^{n} M_{a_{i}|x_{i}}\right].
    \label{probb}
\end{equation}
Here, $M_{a_{i}|x_{i}}$ denotes the positive operator-valued measure (POVM) element associated with outcome $a_{i}$ for measurement setting $x_{i}$. It follows directly that, for separable states, Eq.~\eqref{probb} implies Eq.~\eqref{local_description}. Consequently, any nonlocal state must necessarily be entangled. More generally, the set of local states $\mathcal{L}$ may coincide with the intersection of several sets defined for fixed scenarios.
 
As mentioned previously, nonlocality expresses the incompatibility of probability distributions obtained in quantum mechanics, or more general nonlocal theories, with those allowed by classical theories. This observation leads naturally to a geometric quantification of quantum nonlocality in terms of the trace distance between classical and quantum probability distributions \cite{Brito2018}.
Such a geometrical approach provides a direct characterization of Bell nonlocality and allows to introduce well-behaved \textit{nonlocality monotones}, in close analogy with the resource theory of entanglement.  

In this work, inspired by resource-theoretical approaches~\cite{Chitambar2019, Theurer2017, Gallego2015, Wolfe2020, Dakic2012, Pirandola2014, Streltsov2017, Buscemi2020, Contreras2019, Gour2008, Brandao2013, Buscemi2023, Barrett2005, Schmid2020, Kuroiwa2020, Bartlett2007, DeVicente2014, Gonda2023}, we introduce a natural geometric quantifier of Bell nonlocality based on the distance of a given quantum state to the set of local states. This approach provides a perspective on nonlocality as an intrinsic property of the state and the considered measurement scenario, rather than as a feature associated solely with the resulting probability distribution. In the case where the convex set of local states is defined as the intersection of all possible $\mathcal{L}_{\vec{m}, \vec{\Delta}}$, we obtain geometric measures of nonlocality that depend solely on the quantum state itself.

 We introduce different distance-based monotones, showing that they satisfy the same key properties enjoyed by the \textit{bona fide} entropic and geometric measures of quantum entanglement and other quantum correlations, e.g. quantum discord, introduced in the course of time \cite{Horodecki2009,Plenio2007,Christandl2004,Eisert1999,Wootters1998,Plenio2005,Wei2003,Illuminati2008,Uyanik2010, Geomdiscord2016,Geomdiscord2017,Vedral1997,Vedral1998,Baumgratz2014,Bengtsson2017}. We prove that for Werner states the closest local state is always a Werner state, and analogously, for isotropic states the closest local state is itself isotropic. In the two-qubit case, we also show that the closest local state to a Bell-diagonal state is Bell-diagonal. These results hold independently of the specific Bell inequality under consideration, and therefore capture intrinsic geometric properties of these families of states. 

In practice, the set of local states is known only for specific scenarios. In such cases, the resulting geometric measure not only quantifies nonlocality within the given scenario but also provides a lower bound for more general ones.

The work is organized as follows: In Sect. \ref{sec:free_operations}, we introduce the set of \textit{free operations} as the set of operations that map the set of local states into itself. In Sect. \ref{sec:Geometric_and_entropic_distances_and_their_properties} we introduce the geometric and entropic measures of nonlocality and discuss their properties. In Sect. \ref{sec:two_qubit_case} we address the specific case of two qubit-systems, focusing in Sect. \ref{sec:geometric_measures} on the CHSH inequality. In Sect. \ref{sec:cglmp_inequality} we discuss isotropic and Werner states in arbitrary dimensions, including both bipartite and multipartite systems. As relevant examples, we compute the exact expressions of the geometric measures of nonlocality relative to the bipartite Collins–Gisin–Linden–Massar–Popescu (CGLMP) inequality for two-qudit isotropic states and we present a preliminary discussion of the multipartite Mermin inequality in arbitrary dimensions. Finally, in section \ref{sec:conclusions} we review our results and comment on some possible upcoming research directions in the geometric theory of quantum nonlocality.

\section{Bell nonlocality: the free operations} \label{sec:free_operations}

The set of local states will be denoted in the text as $\mathcal{L}$. In the framework of resource theory, local states are also referred to as free states, meaning that they do not possess any resource -- in this case, nonlocality.  

The set of operations that map local states into local states, also known as the set of free operations, consists of local operations and shared randomness (LOSR) \cite{Chitambar2019,Schmid2020}. These operations can be generally defined as follows:  

\begin{equation}
\int_\lambda d\lambda \Phi^A_\lambda\otimes\Phi^B_\lambda,
\end{equation}

where $\Phi^A_\lambda$ and $\Phi^B_\lambda$ are completely positive and trace-preserving (CPTP) maps. These transformations cannot take a local state outside the set of local states, as they preserve the hidden variable model. Since LOSR operations are also LOCC (while the converse is not true), they inherit many properties established for the extensively studied class of LOCC transformations. There are a few general properties that free operations should satisfy \cite{Chitambar2019}:

\begin{itemize}
\item Free operations preserve the set of free states.
\item Their concatenation is still a free operation.
\item The trivial extension of a free operation is also a free operation (as a consequence, tensor products of free operations are free operations).
\item The operation of adding an auxiliary system is free.
\item The operation of discarding a subsystem (partial trace) is free.
\end{itemize}
In particular, LOSR satisfy all the properties listed above.

\section{Geometric and Entropic Measures of Bell nonlocality}
\label{sec:Geometric_and_entropic_distances_and_their_properties}

\subsection{Geometric measures}

For a given quantum state $\rho$ we take as measure $\mathcal{M}(\rho)$ of Bell nonlocality any contractive distance (with the exception of the Hilbert-Schmidt distance) $D$ between $\rho$ and the closest local state $\rho^{loc}$:
\begin{equation}
\label{general_measure_nl}
\mathcal{M}(\rho) = \min_{\rho^{loc} \in \mathcal L}D(\rho,\rho^{loc}) \, ,
\end{equation}
where $\mathcal{L} \equiv \{ \rho^{loc} \}$ denotes the set of local states.

Physically meaningful and mathematically sound geometric measures of a quantum resource should satisfy the following key properties \cite{Plenio2007,Christandl2004,Eisert1999,Wootters1998,Plenio2005,Wei2003,Illuminati2008,Uyanik2010, Geomdiscord2016,Geomdiscord2017,Vedral1997,Vedral1998,Baumgratz2014,Bengtsson2017}:

\begin{itemize}
\item Property 1 -- vanishing on local states. By definition, this property is automatically satisfied by the relative entropy and by any chosen distance. 
\item Property 2 -- monotonicity under LOSR. For any distance $D$ that is monotonically nonincreasing under an arbitrary CPTP map $\Phi$ that preserves the locality of a local state (LOSR do satisfy this property) we have
\begin{eqnarray}
\mathcal{M}(\rho) &=& \min_{\rho^{loc} \in \mathcal L} D(\rho,\rho^{loc}) = D(\rho,\rho^{loc}_{min}) \nonumber \\
&\geq& D(\Phi(\rho),\Phi(\rho^{loc}_{min})) \geq \min_{\rho^{loc} \in \mathcal L} D(\Phi(\rho),\rho^{loc}) = \mathcal{M}(\Phi(\rho)) .
\end{eqnarray}

It follows that any distance that is nonincreasing under CPTP maps defines a measure of nonlocality that is nonincreasing under LOSR. Property 2 holds for the relative entropy as well, since the latter is contractive under CPTP maps by construction.
\item Property 3 -- convexity (nonincreasing under mixing of quantum states). Since the set of local states is convex any distance that is jointly convex induces a measure of nonlocality that is convex. Denoting by $a$ the convexity parameter with $0 \leq a \leq1$, we have:
\begin{eqnarray}
\hspace{-2cm}a \mathcal{M}(\rho_1)+(1-a)\mathcal{M}(\rho_2)&=&a\min_{\rho^{loc} \in \mathcal L} D(\rho_1,\rho^{loc})+(1-a)\min_{\rho^{loc} \in \mathcal L}D(\rho_2,\rho^{loc}) \nonumber\\
&=& a D(\rho_1,\rho^{loc}_{min})+(1-a) D(\rho_2,\rho^{'loc}_{min}) \nonumber \\
&\geq& D(a\rho_1+(1-a)\rho_2,a\rho^{loc}_{min}+(1-a)\rho^{'loc}_{min}) \nonumber\\
&\geq& \min_{\rho^{loc} \in \mathcal L}D(a\rho_1+(1-a)\rho_2,\rho^{loc}) \nonumber \\
&=& \mathcal{M}(a\rho_1+(1-a)\rho_2).
\end{eqnarray}
Joint convexity holds for the relative entropy by definition, and thus the latter is automatically nonincreasing under mixing.
\item Property 4 -- monotonicity under subselection. We require monotonicity under selective measurements on average:
\begin{equation}
\mathcal{M}(\rho) \geq \sum_{i} p_i \mathcal{M}\left(\frac{\Lambda_i \rho \Lambda_i^{\dagger}}{p_i} \right) ,
\end{equation}
where the weights $p_i = {\rm Tr}(\Lambda_i \rho \Lambda_i^{\dagger})$ and the set $\{\Lambda_i\}$ of Kraus operators is such that $\sum_{i}\Lambda_i^{\dagger}\Lambda_i = \mathds{1}$ and $\Lambda_i \rho \Lambda_i^{\dagger} \in \mathcal L$ $\forall \rho^{loc} \in \mathcal L$. In a sense, property 4 might be considered even more important than property 2, since subselection is a process often readily available under controlled experimental conditions. On the other hand, the former is usually significantly harder to verify than the latter, which is certainly satisfied by all well behaved geometric and entropic measures. It should be noted that convexity (property 3) and monotonicity under subselection (property 4), automatically imply monotonicity under LOSR (property 2). The reverse is obviously not true in general. Focusing only on the relative entropy, proof of property 4 is quite straightforward, in analogy with what has been shown to hold for the resource theory of entanglement \cite{Vedral1998} and quantum coherence \cite{Baumgratz2014}, and we provide it in Section \ref{sec:rel_entropy}.
\item Property 5 -- computability. This property depends on the distance. We will study this subject case by case in the following by computing explicitly various measures for selected classes of two-qubit states.
\end{itemize}

In the following, we introduce geometric and entropic measurements and analyze the properties they satisfy.

\subsubsection{Hilbert-Schmidt distance}
The Hilbert-Schmidt distance $D_\mathrm{HS}(\rho_{1} , \rho_{2})$ between two density operators $\rho_{1}$ and $\rho_{2}$ is defined through the Hilbert-Schmidt norm as follows:
\begin{equation}
D_\mathrm{HS}(\rho_{1} , \rho_{2}) \, = \, \|\rho_1-\rho_2\|_{2} \, = \, \sqrt{{\rm Tr} \left[ (\rho_1 - \rho_2)(\rho_1 - \rho_2) \right] } \, \, .
\label{H-S_distance}
\end{equation}
 and the corresponding measure is
\begin{equation}
\mathcal{M}_{\mathrm{HS}}(\rho) \, = \, \min_{\rho^{loc} \in \mathcal L} D_{\mathrm{HS}}(\rho,\rho^{loc}) \, .
\end{equation}
The Hilbert-Schmidt (HS) distance has no direct operational meaning and, in general, may not monotonic under all CPTP transformations \cite{Bengtsson2017, Piani2012}. For instance, it increases under a trivial extension of local subsystems. On the other hand, it provides easily computable bounds on valid contractive distances that are often more difficult to compute.

\subsubsection{Hellinger distance}
The Hellinger distance is defined as the Euclidean two-norm of the difference of the square roots of the density matrices (taking squares for computational convenience):
\begin{equation}
\label{Hellinger_distance}
    D_\mathrm{He}(\rho_1, \rho_2) = \|\sqrt{\rho_1}-\sqrt{\rho_2}\|_2 .
\end{equation}
Given a pair of states that are diagonal in the same basis with, respectively, eigenvalues $\{e^1_i\}$ and $\{e^2_i\}$, one has: 
\begin{equation}
    D_\mathrm{He}(\rho_1,\rho_2)=\sqrt{2-2\sum_i \sqrt{e^{(1)}_ie^{(2)}_i}} = \sqrt{\displaystyle\sum_{i=1}^{4} \bigg( \sqrt{e_{i}^{(1)}}-\sqrt{e_{i}^{(2)}} \bigg)^{2}} .
    \label{HellingerDistance}
\end{equation}
The above is the Euclidean norm in the $\sqrt{e_{i}}-$space and coincides with the classical Hellinger distance quantifying the distance between the two classical probability distributions \cite{Ref_Hellinger1909}. The corresponding Hellinger measure of Bell nonlocality is thus defined as
\begin{equation}
    \mathcal{M}_{\mathrm{He}}(\rho) = \min_{\rho^{loc} \in \mathcal L} D_{\mathrm{He}}(\rho,\rho^{loc}) \, .
\end{equation}

\subsubsection{Bures distance}
The Bures distance is defined as \cite{Bengtsson2017}
\begin{equation}
    D_\mathrm{B}(\rho_1,\rho_2) = \sqrt{2\big[ 1-\sqrt{F(\rho_{1},\rho_{2}}) \big]} \, ,
\end{equation}
where $F$ is the Uhlmann fidelity given by \cite{Uhlmann1976,Nielsen2005}
\begin{equation}
F(\rho_{1},\rho_{2}) = \bigg(\mathrm{Tr}\bigg[ \sqrt{\sqrt{\rho_{1}}\rho_{2}\sqrt{\rho_{1}}} \bigg] \bigg)^{2} \, .
\end{equation}
In the case of commuting density operators, the Bures and Hellinger distances coincide. The corresponding Bures measure of Bell nonlocality reads
\begin{equation}
\mathcal{M}_{\mathrm{B}}(\rho) = \min_{\rho^{loc} \in \mathcal L} D_{\mathrm{B}}(\rho,\rho^{loc}) \, .
\end{equation}

\subsubsection{Trace distance}
The trace distance is defined as
\begin{equation}
    D_\mathrm{Tr}(\rho_1,\rho_2)= \frac{1}{2} \|\rho_1-\rho_2\|_1 = \frac{1}{2} {\rm Tr} \left[ \sqrt{(\rho_1 - \rho_2)(\rho_1 - \rho_2)} \right] \, ,
    \label{trace_distance}
\end{equation}
where $\| \mathcal{O} \|_1 \equiv \sqrt{\mathcal{O}^{\dagger}\mathcal{O}}$. For states that are diagonal in the same basis with eigenvalues 
$\{e^1_i\}$ and $\{e^2_i\}$ respectively, we have
\begin{equation}
    D_\mathrm{Tr}(\rho_1,\rho_2)= \frac{1}{2} {\rm Tr} | \rho_1 - \rho_2 | = \frac{1}{2} \sum_i|e_i^1-e^2_i| \, .
\end{equation}
The corresponding trace-distance measure of Bell nonlocality reads
\begin{equation}
    \mathcal{M}_\mathrm{Tr}(\rho) = \min_{\rho^{loc} \in \mathcal L} D_\mathrm{Tr}(\rho,\rho^{loc}) .
    \label{tracemeasure}
\end{equation}
The trace distance defines the probability of error in the Helstrom measurement which is the optimal measurement to distinguish two quantum states. It is jointly convex and nonincreasing under CPTP maps \cite{Bengtsson2017,Nielsen2005}. As a consequence, the corresponding measure of Bell nonlocality Eq.~\eqref{tracemeasure} satisfies monotonicity under LOSR and monotonicity under mixing (convexity).

\subsection{Relative entropy}
\label{sec:rel_entropy}
Relative entropy is not a distance as it does not satisfy the triangle inequality and is not symmetric with respect to state interchange, but nonetheless it is widely used in the quantum information theory and the theory of entanglement because it enjoys nearly all of the monotonicity properties that are requested in a valid quantum resource theory. The relative entropy is defined as
\begin{equation}
    S(\rho_1||\rho_2) = {\rm Tr}\rho_1 \log_2\rho_1 - {\rm Tr}\rho_1 \log_2\rho_2. 
    \label{relativeEntropygeneral}
\end{equation}
For states that are diagonal in the same basis with eigenvalues $e^1_i$ and $e^2_i$ respectively we have 
\begin{equation}
    S(\rho_1||\rho_2) = \sum_ie^1_i \log_2e^1_i - \sum_ie^1_i \log_2e^2_i. 
    \label{relativeEntropy}
\end{equation}
Hence, we define the measure of Bell nonlocality based on the relative entropy as
\begin{equation}
    \mathcal{M}_{\mathrm{Re}}(\rho) = \min_{\rho^{loc} \in \mathcal L} S(\rho || \rho^{loc}) .
    \label{relativeEntropymeasure}
\end{equation}
We will show here that the measure defined by Eq.~\eqref{relativeEntropymeasure} fulfills monotonicity under subselection, i.e.:
\begin{eqnarray}
    \mathcal{M}_{\mathrm{Re}}(\rho) \geq \displaystyle\sum_{i}p_{i}\mathcal{M}_{\mathrm{Re}} (\rho_{i}) ,
\end{eqnarray}
where $\{ \Lambda{i} \}$ is any set of Kraus operators, with weights $p_i = {\rm Tr}(\Lambda_i \rho \Lambda_i^{\dagger})$, under the hypothesis that the involved Kraus operators are free, i.e. preserve the local states: $\sum_{i}\Lambda_i^{\dagger}\Lambda_i = \mathds{1}$ and $\Lambda_i \rho \Lambda_i^{\dagger} \in \mathcal{L}$ $\forall \rho^{loc} \in \mathcal{L}$, where $\mathcal{L}$ is the set of local states.

To obtain the last inequality, first of all note that since the relative entropy is invariant under the same unitary or isometry applied to both arguments, we can use the isometry $I=\sum_i|i\rangle_{aux}\otimes|i\rangle_{aux'}\otimes V_i$ for any $V_i$ to derive the identity $S(\rho^{(1)}|\rho^{(2)}) = S(I\rho^{(1)} I^\dagger|I\rho^{(2)} I^\dagger)$, for any couple of states $\rho^{(1)}$ and $\rho^{(2)}$, where $\{|i\rangle_{aux}\}_i$ is an orthogonal basis of an auxiliary system. Next, monotonicity of the relative entropy under the partial trace over the auxiliary system $aux$ implies  
\begin{equation}
S(\rho^{(1)}||\rho^{(2)})\geq \sum_i S(\tilde{\rho}_i^{(1)}||\tilde{\rho}_i^{(2)}),
\end{equation}
where we use unnormalized states $\tilde{\rho}_i^{(1)} = V_i\rho^{(1)} V_i^{\dagger}$ and $\tilde{\rho}_i^{(2)} = V_i\rho^{(2)} V_i^{\dagger}$.  
Following the argument derived in \cite{Vedral1998} we have 
\begin{equation}
\sum_i S(\rho_i^{(1)}||\rho_i^{(2)})\geq \sum_i p_i S\left(\frac{\tilde{\rho}_i^{(1)}}{p_i}\bigg|\frac{\tilde{\rho}_i^{(2)}}{q_i}\right),
\end{equation}
where $p_i={\rm Tr}\ \tilde{\rho}_i^{(1)}$ and $q_i={\rm Tr}\ \tilde{\rho}_i^{(2)}$.
Identifying $\rho \equiv \rho^{(1)}$, assuming that $\rho^{loc} \equiv \rho^{(2)}$ is a local state and $V_i$ are Kraus operators which preserve locality, such that $V_i\rho^{loc} V_i^{\dagger}/q_i$ is still a local state, we conclude that
\begin{equation}
\mathcal{M}_{\mathrm{Re}}(\rho)\geq\sum_i p_i\mathcal{M}_{\mathrm{Re}}\left(\frac{V_i\rho V_i^{\dagger}}{p_i}\right).
\end{equation}

\section{Application to two-qubit states} \label{sec:two_qubit_case}
 An arbitrary two-qubit state $\rho$ can be represented as
\begin{equation}
    \rho = \frac{1}{4}\sum_{i,j = 0}^3 \alpha_{ij} \, \sigma_i \otimes \sigma_j,
    \label{representation}
\end{equation}
where $\sigma_0 = \mathds{1}_2$ and $\sigma_i$ ($i=1,2,3$) denote the Pauli matrices, satisfying ${\rm Tr}[\sigma_i] = 0$ and ${\rm Tr}[\sigma_i\sigma_j] = 2\delta_{ij}$. The expansion coefficients are given by $\alpha_{00} = 1$, $\alpha_{0i} = {\rm Tr}\big[ \rho \, ( \mathds{1}_2 \otimes \sigma_i ) \big]$, $\alpha_{i0} = {\rm Tr}\big[ \rho \, ( \sigma_i \otimes \mathds{1}_2 ) \big]$, and $\alpha_{ij} = {\rm Tr}\big[ \rho \, ( \sigma_i \otimes \sigma_j ) \big]$ for $i=1,2,3$. These coefficients can be arranged in matrix form as
\begin{equation}
    \alpha = 
    \left[
    \begin{array}{cc}
    1 & \vec{\alpha}_{0i} \\ 
    \vec{\alpha}_{i0} & A
    \end{array}
    \right],
\label{alpha_matrix_generic}
\end{equation}
where all entries are real, $\vec{\alpha}_{0i}$ is a row vector with components $\{\alpha_{0i}\}$, $\vec{\alpha}_{i0}$ is a column vector with components $\{\alpha_{i0}\}$, and $A$ is a $3 \times 3$ real matrix. As first result, we prove the following: \\

{\bf Result 1.} {\it If $\rho$ is a local state with given $(\alpha_{0i}, \alpha_{i0}, A)$ then the state $\rho^-$ given by 
$(-\alpha_{0i}, -\alpha_{i0}, A)$ is also local.}
\vspace{0.2cm}
\begin{myproof}
It suffices to consider a local transformation that for $i=1,2,3$ changes $\sigma_i$ into $-\sigma_i$.
\end{myproof}
\vspace{0.3cm}
{\bf Result 2.} {\it For any local state $\rho$ with given $(\alpha_{0i}, \alpha_{i0}, A)$ there exists a local state $\rho'$ with the same $A$ but  with maximally mixed reduced states, i.e., with $\alpha_{0i}=\alpha_{i0}=0$}
\vspace{0.2cm}
\begin{myproof}
    Consider the convex combination 
\begin{equation}
\rho' = \frac{1}{2}\rho + \frac{1}{2}\rho^-.
\end{equation}
Indeed, for such a combination we have
\begin{equation}
    \alpha = 
    \left[
    \begin{array}{cc}
    \alpha_{00} & 0 \\ 
    0 & A
    \end{array}
    \right].
\end{equation}
Now, convex combinations of local states are local and this ends the proof.
\end{myproof}
\vspace{0.2cm}
Two particularly relevant families of states are the Bell-diagonal states, i.e., convex mixtures of the four Bell states (see~\ref{appendix_1} for details):
\begin{eqnarray}
    \rho_\mathrm{BD} = \sum_{i=1}^{4} e_{i} \ket{\Psi_{i}}\bra{\Psi_{i}} 
    = \bigg[ \mathds{1}_{2} \otimes \mathds{1}_{2} + \sum_{i=1}^{3} a_{i} \, \sigma_{i} \otimes \sigma_{i} \bigg] \, ,
    \label{Bell-diagonal states in termini delle sigma}
\end{eqnarray}
and the subclass known as Werner states, which are convex combinations of the maximally mixed state and the singlet Bell state. They are parametrized by a single parameter $\omega$ and were introduced as the first example of states that are entangled yet do not necessarily violate the CHSH inequality~\cite{Werner1989}.

For this class of states we prove the following two propositions, holding for every scenario in which a set of Bell inequalities (just ``Bell inequality" in the following) defines the convex set $\mathcal L$ of local states in that scenario:

\noindent {\bf Proposition 1.} {\it For any Bell inequality, any jointly convex and unitarily invariant functional $D$ of pairs of two-qubit density matrices}
\begin{equation}
    \min_{\rho^{loc} \in \mathcal L}D(\rho_{w},\rho^{loc}) =  \min_{\rho^{loc}_w \in \mathcal L}D(\rho_{w},\rho^{loc}_w) \, ,
\label{Proposition1}
\end{equation}
{\it where $\rho_w$ is a Werner state, $\rho^{loc}$ is a local state, and $\rho^{loc}_w$ is a local Werner state.\\}

\begin{myproof}
The proof of Proposition 1 is shown in section~\ref{sec:cglmp_inequality} for isotropic states and generalized Werner states in higher dimensions.
Here we just sketch the relevant steps. Starting from Eq.~\eqref{general_measure_nl}, we can apply the same $U\otimes U$ to both arguments of $D$ and make use of the unitarily invariance of $D$; then, we bound this quantity from below by twirling both arguments and using jointly convexity; finally we recognize that twirling  projects $\rho_w$  itself and $\rho^{loc}$ into another (local) Werner state $\rho^{loc}_w$. So $D(\rho_w,\rho^{loc}_w)\leq D(\rho_w,\rho^{loc})$, hence the closest state can be taken to be a Werner.
\end{myproof}

\vspace{0.2cm}

\noindent {\bf Proposition 2.} {\it For any Bell inequality, any jointly convex and unitarily invariant functional $D$ of pairs of two-qubit density matrices, the closest local state to a Bell-diagonal state is Bell-diagonal: 
\begin{equation}
\min_{\rho^{loc} \in \mathcal L}D(\rho_{\mathrm{BD}}, \rho^{loc}) = \min_{\rho_{\mathrm{BD}}^{loc} \in \mathcal L}D(\rho_{\mathrm{BD}}, \rho_{\mathrm{BD}}^{loc}) \, ,
\label{Proposition2}
\end{equation}
where $\rho_{\mathrm{BD}}$ denotes a Bell-diagonal state, $\rho^{loc}$ a local state, and $\rho_{\mathrm{BD}}^{loc}$ a local Bell-diagonal state.
}\\ 

\begin{myproof}

The symmetry class that we consider contains the simultaneous local $\pi$-rotations of qubits around the three Pauli axes. Rotation $R_1$ around $\sigma_1$ changes $\sigma_2$ and $\sigma_3$ to $-\sigma_2$ and $-\sigma_3$, respectively. Rotations $R_2$ and $R_3$ operate analogously. These transformations are local and unitary, therefore, they cannot change the locality of a quantum state. Moreover, these transformations preserve Bell-diagonal states. Indeed, let us consider the general state representation Eq. (\ref{representation}). Under rotation $R_1$, we have the following transformation:
\begin{equation}
\hspace{-1cm}
\alpha =  
\left(
\begin{array}{cccc}
1 & a_{01} & a_{02} & a_{03} \\
a_{10} & a_{11} & a_{12} & a_{13} \\
a_{20} & a_{21} & a_{22} & a_{23} \\
a_{30} & a_{31} & a_{32} & a_{33}
\end{array}
\right)
\, \, \longrightarrow \, \,
\alpha_1 =  
\left(
\begin{array}{cccc}
1 & a_{01} & -a_{02} & -a_{03} \\
a_{10} & a_{11} & -a_{12} & -a_{13} \\
- a_{20} & - a_{21} & a_{22} & a_{23} \\
- a_{30} & - a_{31} & a_{32} & a_{33}
\end{array}
\right).
\end{equation}

The average $\rho'=1/2(\rho + R_1(\rho))$ enjoys the representation
\begin{equation}
\hspace{-1cm}
\alpha_1' =  
\left(
\begin{array}{cccc}
1 & a_{01} & 0 & 0 \\
a_{10} & a_{11} & 0 & 0 \\
0 & 0 & a_{22} & a_{23} \\
0 & 0 & a_{32} & a_{33}
\end{array}
\right).
\end{equation}

Applying rotation $R_2$ to $\rho'$ and averaging the result with $\rho'$ leads to the state with diagonal representation Eq. (\ref{Bell-diagonal states in termini delle sigma}), i.e., a Bell-diagonal state. The total transformation that projects a given state to a Bell-diagonal state takes the compact form
\begin{equation}
\hspace{-1cm} \rho \, \, \longrightarrow \, \, \frac{1}{2}\left(\frac{1}{2}(\rho + R_1(\rho))+R_2\left(\frac{1}{2}(\rho + R_1(\rho))\right)\right).
\end{equation}

This transformation obviously preserves the Bell-diagonal states, does not change the locality, and is a composition of joint local unitaries and averaging.

The proof of Proposition 2 is thus immediately completed by straightforwardly adapting the reasoning followed in the proof of Proposition 3.
\end{myproof} 

\vspace{0.2cm}

Note that the previous result is useful for quantifying nonlocality of any two-qubit system with maximally-mixed subsystems, since those states can be transformed in Bell-diagonal states by means of local unitary transformations which do not change the nonlocality of the state.

Indeed, if one considers the singular value decomposition of the matrix $A$:
\begin{equation}
    A = O^1DO^2,
\label{singular}    
\end{equation}
where $O^1$ and $O^2$ are orthogonal matrices and $D$ is diagonal and non-negative. The orthogonal transformations correspond to local changes of basis:  
\begin{eqnarray}
    \rho = \frac{1}{4}&\bigg(&\mathds{1}_{4}+\sum_{i=1}^3\alpha_{0i}\mathds{1}_{2}\otimes \sigma_i \nonumber  + \sum_{i=1}^3\alpha_{i0} \sigma_i\otimes \mathds{1}_{2} \nonumber \\ &+&\sum_{k=1}^3D_{k}\bigg(\sum_iO^1_{ik}\sigma_i\bigg)\otimes\bigg(\sum_jO^2_{kj}\sigma_j\bigg)\bigg).
\end{eqnarray}

\section{CHSH-nonlocality of two-qubit states}\label{sec:geometric_measures}

Exploiting the results derived in the previous section, we can proceed to quantify the CHSH nonlocality of two-qubit states using the Hilbert space based distance and relative entropy measures with respect to the set of CHSH-local two-qubit states. 

In the case of two dichotomic observables per observer ($m=2$, $\Delta=2$), the Bell inequality takes the CHSH form:  

\begin{equation}
E(x,y)-E(x,y')+E(x',y)+E(x',y')\leq 2,
\label{CHSH}
\end{equation}
where $(x,x')$ and $(y,y')$ are the observables of Alice and Bob, respectively. $E(x,y)$ denotes the statistical expectation value of the measurement outcomes for observables $x$ and $y$ in a given quantum state. For nonlocal states, Eq.~\eqref{probb} does not hold, leading to a violation of the CHSH inequality, Eq.~\eqref{CHSH}.

Let $(d_1, d_2, d_3)$ be the singular values in the decomposition Eq.~\eqref{singular}, and without loss of generality, let $d_1\geq d_2\geq d_3$. As proved in Ref. \cite{Horodecki1995}, a state is CHSH-local, i.e., it does not violate the CHSH inequality Eq.~\eqref{CHSH}, if and only if
\begin{equation}
    d_1^2+d_2^2\leq 1.
    \label{local}
\end{equation}

Explicit examples will include Werner states, Bell-diagonal states, and reduced convex combinations of Bell states. These important classes of states are routinely considered in the study of quantum resource theories, starting from the pioneering investigations on entanglement and discord \cite{Wootters1997, Wootters1998, Henderson_Vedral2001}.

\subsection{Werner states}
\subsubsection{Hilbert-Schmidt distance}

Consider the Hilbert-Schmidt (HS) distance between a generic pair $\{ \rho_1 , \rho_2 \}$ of two-qubit states as induced by the HS norm Eq.~\eqref{H-S_distance}. Using Eq. (\ref{representation}) and the properties of the Pauli matrices:

\begin{equation}
   \hspace{-1cm}  \|\rho_1-\rho_2\|_{2} \, =\frac{1}{2} \, \|\alpha_1-\alpha_2\|_{2} \, .
    \label{Euclid}
\end{equation}
It follows from Proposition 1, Eq.~\eqref{Proposition1}, that the closest CHSH-local state to a Werner state is also a Werner state with $w \leq 1/\sqrt{2}$. Hence, for CHSH-nonlocal Werner states, i.e., with $w \geq 1/\sqrt{2}$ :
\begin{equation}
\hspace{-1cm} \mathcal{M}_\mathrm{HS}(\rho_{w}) = \min_{\rho^{loc}_{w} \in \mathcal{L}}\left\|\rho_w -\rho^{loc}\right\|_{2}  = \frac{\sqrt{3}}{2}\left(w- \frac{1}{\sqrt{2}}\right)\, ,
\label{measurehswerner}
\end{equation}
where $\mathds{1}$ is the identity matrix and $w \in \left(1/\sqrt{2} , 1\right]$. Correctly, the maximum is achieved by the maximally entangled Bell states corresponding to $w=1$ :
\begin{equation}
\mathcal{M}_\mathrm{HS}^{max} \, = \, {\frac{\sqrt 3}{ 2 } \left( 1 -  \frac{1}{\sqrt 2}\right)} \, .  
\label{HSmax}
\end{equation}
The behavior of the normalized HS geometric measure of CHSH nonlocality for Werner states $\widetilde{\mathcal{M}}_\mathrm{HS} = \mathcal{M}_\mathrm{HS}(\rho_{w})/\mathcal{M}_\mathrm{HS}^{max}$ as a function of the $w$ parameter is reported in Fig. \ref{fig:Werner}. The plot is a parabola, as the HS distance is simply the Euclidean distance in the three-dimensional space of Fig. \ref{fig:pic}.

\subsubsection{Hellinger distance}
Recalling again the consequences of Proposition 1, Eq.~\eqref{Proposition1}, the measure of CHSH nonlocality for Werner states based on the (squared) Hellinger distance, Eq.~\eqref{HellingerDistance}, reads
\begin{eqnarray}
   \hspace{-1cm}\mathcal{M}_\mathrm{He}(\rho_{w}) &=& \min_{\rho_w^{loc} \in \mathcal{L}}D^2_\mathrm{He}(\rho_w, \rho_w^{loc}) \nonumber \\
   &=& 2-\frac{1}{2}\left(3\sqrt{(1-w)\left(1-\frac{1}{\sqrt{2}}\right)}+\sqrt{(1+3w)\left(1+\frac{3}{\sqrt{2}}\right)}\right) \, ,
   \label{Werner_Hellinger}
\end{eqnarray}
where $w\in(1/\sqrt{2}, 1]$.
Here we have used the relation between the eigenvalues of the state density matrix and the Werner parameter $w$, which is derived directly from Eq. (\ref{eigenvalues}). As $|a_1|=|a_2|+|a_3|=w$, the signs of the matrix elements $a_i$ identify the states corresponding to different corners of the state tetrahedron in Fig. \ref{fig:pic}. The Hellinger measure is maximized by the Bell states at $w=1$ :
\begin{equation}
\mathcal{M}_\mathrm{He}^{max}=2-\sqrt{1 + \frac{3}{\sqrt{2}}} \, \, .
\end{equation}
The behavior of the normalized Hellinger geometric measure of CHSH nonlocality for Werner states $\widetilde{\mathcal{M}}_\mathrm{He} = \mathcal{M}_\mathrm{He}(\rho_{w})/\mathcal{M}_\mathrm{He}^{max}$ as a function of the $w$ parameter is shown in Fig. \ref{fig:Werner}.

\subsubsection{Bures distance}
In the case of commuting operators, the Bures distance and the Hellinger distance coincide. This is indeed the case for Werner density matrices. Therefore, the Bures and Hellinger measures of CHSH nonlocality coincide on Werner states $\rho_{w}$:
\begin{equation}
\mathcal{M}_\mathrm{Bu}(\rho_{w}) = \mathcal{M}_\mathrm{He}(\rho_{w}) \, .  
\end{equation}

\subsubsection{Trace distance}
The geometric measure of CHSH nonlocality of Werner states based on the trace distance Eq.~\eqref{trace_distance} reads:
\begin{equation}
    \mathcal{M}_\mathrm{Tr}(\rho_{w}) = \min_{\rho_w^{loc} \in \mathcal{L}}D_\mathrm{Tr}(\rho_w, \rho_w^{loc}) = \frac{3}{4}\left|w-\frac{1}{\sqrt{2}}\right| = \frac{3}{4}\left( w-\frac{1}{\sqrt{2}} \right) \, ,
\end{equation}
where the last equality follows from the fact that, without loss of generality, we can limit the analysis to the interval $w\in(1/\sqrt{2}, 1]$. The maximum value is achieved at $w=1$ by the Bell states: 
\begin{equation}
 \mathcal{M}_\mathrm{Tr}^{max} = \frac{3}{4}\left(1-\frac{1}{\sqrt{2}}\right) \, .   
\end{equation}
 The normalized trace measure of CHSH nonlocality for Werner states $\widetilde{\mathcal{M}}_\mathrm{Tr} = \mathcal{M}_\mathrm{Tr}(\rho_{w})/\mathcal{M}_\mathrm{Tr}^{max}$ as a function of the $w$ parameter is shown in figure \ref{fig:Werner}.

\subsubsection{Relative entropy}
It is straightforward to verify that for a nonlocal Werner state in the interval $w \geq 1/\sqrt{2}$, its relative entropy Eq.~\eqref{relativeEntropy} with respect to the set $\sigma$ of local states is minimized by the local state with $w = 1/\sqrt{2}$ , so that:
\begin{eqnarray}
\hspace{-2cm} \mathcal{M}_{\mathrm{Re}}(\rho_w) = \min_{\rho_w^{loc} \in \mathcal{L}}S(\rho_w | \rho_w^{loc}) & = & \frac{3(1-w)}{4} \log_2 \left( \frac{3(1-w)}{4} \right) + \frac{1+3w}{4} \log_2 \left( \frac{1+3w}{4} \right) + \nonumber \\ 
&& \nonumber \\ \hspace{-1cm}
& - & \frac{3(1-w)}{4} \log_2 \left( \frac{\sqrt{2} - 1}{4\sqrt{2}} \right) - \frac{1+3w}{4} \log_2 \left( 
\frac{\sqrt{2} + 3}{4\sqrt{2}} \right) \, .   
\end{eqnarray}

\begin{figure}[h]
    \centering
    \includegraphics[scale =.4]{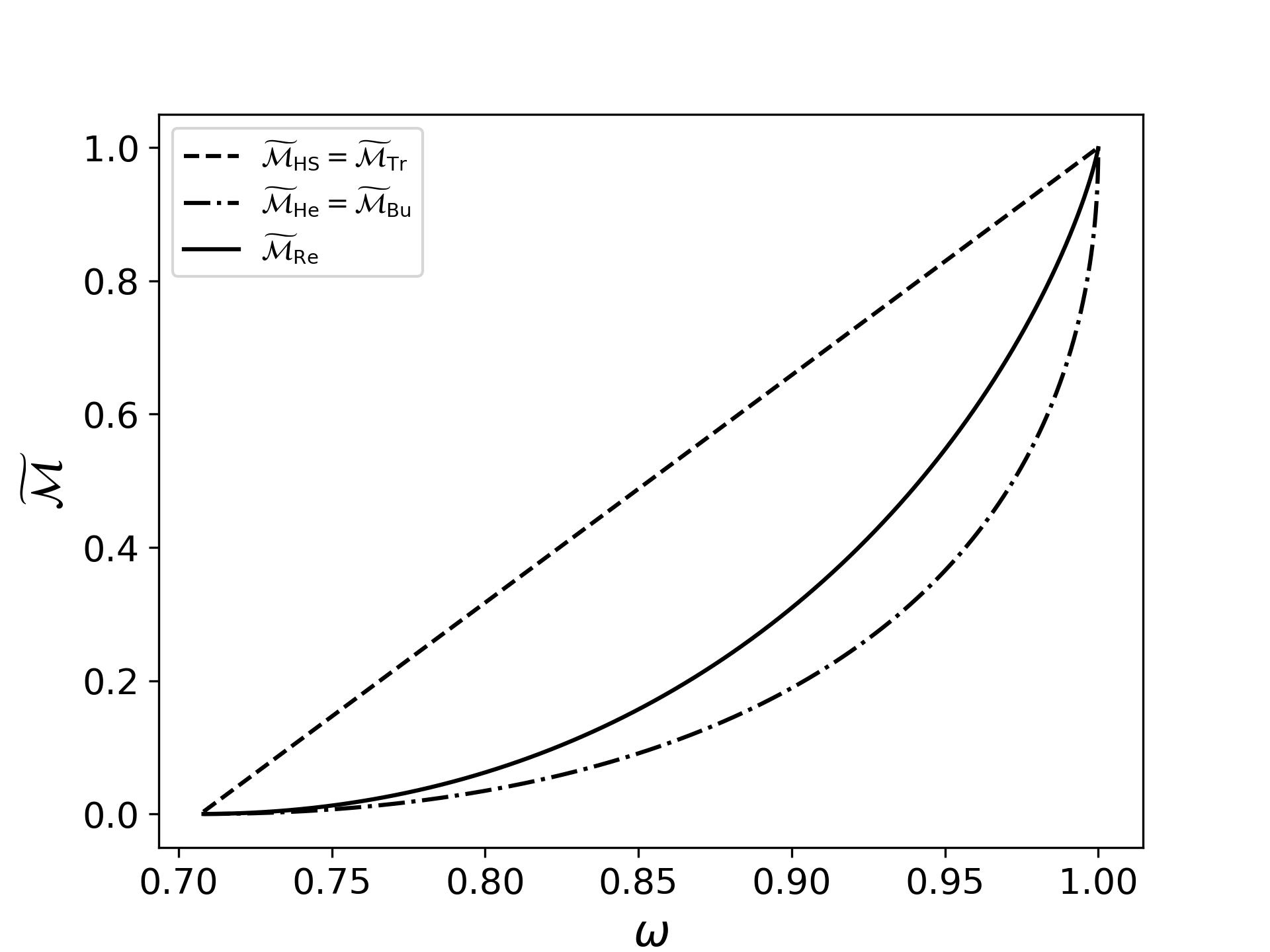}
    \caption{Normalized geometric measures of CHSH nonlocality for Werner states as a function of the Werner parameter $w$ in the interval $\frac{1}{\sqrt{2}} \leq w \leq 1$. Dashed straight line: Trace and Hilbert-Schmidt measures $\widetilde{\mathcal{M}}_\mathrm{Tr} = \widetilde{\mathcal{M}}_{\mathrm{HS}}$. Dashed-dotted curve: Hellinger and Bures measure $\widetilde{\mathcal{M}}_\mathrm{He} = \widetilde{\mathcal{M}}_\mathrm{Bu}$. Solid curve: Relative entropy measure $\widetilde{\mathcal{M}}_{\mathrm{Re}}$. Dashed-dotted curve: Hellinger measure $\widetilde{\mathcal{M}}_\mathrm{He}$. All plotted quantities are dimensionless.}
    \label{fig:Werner}
\end{figure}

The maximum value achieved by the Bell states at $w=1$ reads
\begin{equation}
\mathcal{M}_{\mathrm{Re}}^{max} = 2 - \log_{2}{\bigg(1+\frac{3}{\sqrt{2}}\bigg)} \, .
\end{equation}
The normalized relative entropy measure of CHSH nonlocality for Werner states 
$\widetilde{\mathcal{M}}_{\mathrm{Re}} = \mathcal{M}_{\mathrm{Re}}(\rho_{w})/\mathcal{M}_{\mathrm{Re}}^{max}$ as a function of the $w$ parameter is reported in Fig. \ref{fig:Werner}.

\subsection{Bell-diagonal states}

As mentioned before, the results found in this section can be applied to any two-qubit state with maximally disordered subsystems, since they can be transformed into Bell-diagonal states by local unitary transformations that do not change their nonlocality.

Since Bell-diagonal states are diagonal in the Bell basis, the evaluation of the previously introduced distances is straightforward for this class of states. For a given state $\rho$, the closest local state can be identified through numerical analysis by exploiting the Horodecki criterion,
\begin{equation}
    \max_{i,j \, , \, i\neq j} \{ a_{i}^{2}+a_{j}^{2}\} = 1 \, ,
    \label{Surface_HS}
\end{equation}
(see~\ref{appendix_2} for details). Here, let us consider Bell-diagonal states that are convex combinations of only two of the four Bell states. For instance, let us choose $e_{2}=e_{4}=0$, $e_{1} = p$, and $e_{3}=1-p$, with $p$ as a free parameter in the interval $[0,1]$. Such Bell-diagonal states represent all possible convex combinations of the first and third Bell states of Eq.~\eqref{Bellstates}. Moving to symplectic space and recalling Eqs. \eqref{eigenvalues}, this instance corresponds to $a_{1}=a$, $a_{2}=-a$, and $a_{3}=1$, with $a=2p-1$.

\begin{figure}[H]
    \centering
    \includegraphics[scale =.3]{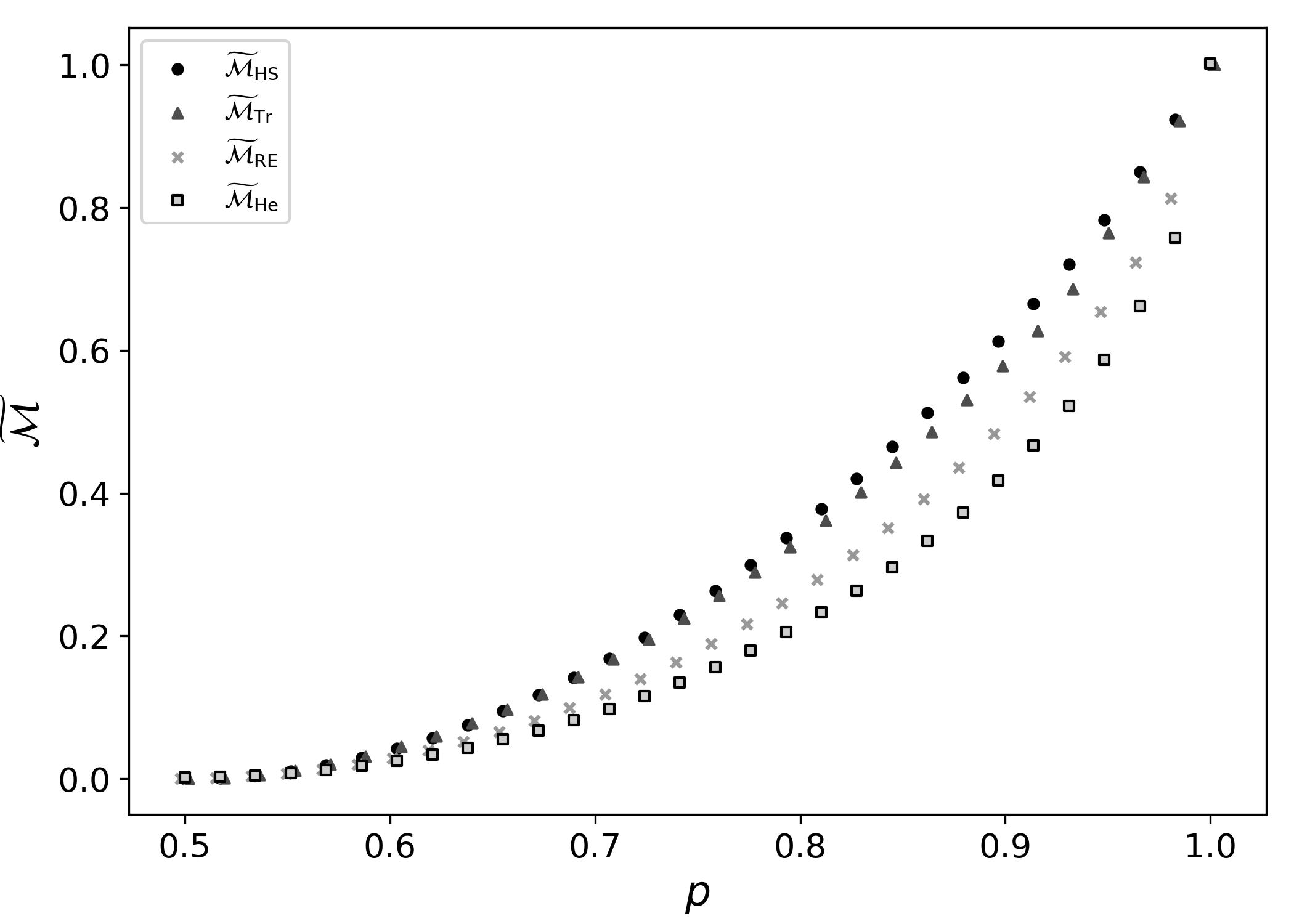}
    \caption{Normalized measures of CHSH nonlocality for the convex combination of two Bell states as functions of the convex combination parameter $p$ in the interval $p \geq \frac{1}{2}$. The same behavior holds in the symmetric case $p \leq \frac{1}{2}$, with the maximum achieved in $p=0$. Black dashed curve: Trace measure $\widetilde{\mathcal{M}}_\mathrm{Tr}$. Black solid curve: Relative entropy measure $\widetilde{\mathcal{M}}_{\mathrm{Re}}$. Grey dashed curve: Hellinger measure $\widetilde{\mathcal{M}}_\mathrm{He}$.  Grey solid curve: Hilbert-Schmidt measure $\widetilde{\mathcal{M}}_\mathrm{HS}$. All plotted quantities are dimensionless.}
    \label{fig:BD_one_parameter}
\end{figure}

For this class of states, the locality condition Eq.~\eqref{local} is fulfilled only for $p=\frac{1}{2}$; therefore, such states are nonlocal for $0\leq p \leq 1$, with the exception of $p=\frac{1}{2}$. The behavior of the four normalized measures of CHSH nonlocality as functions of the convex combination parameter $p$ in the half-interval 
$p \geq \frac{1}{2}$ is reported in Fig. \ref{fig:BD_one_parameter}. We see that all measures vanish at 
$p=\frac{1}{2}$ and reach their maximum at $p=1$, i.e., on the first Bell state. For all measures, the normalization coefficients turn out to be the same as the ones of the nonlocality measures for Werner states, in agreement with Proposition 1, Eq.~\eqref{Proposition1}, and Proposition 2, Eq.~\eqref{Proposition2}. The case $p \leq \frac{1}{2}$ is readily obtained by symmetry. 

It is straightforward to verify that the above-described behavior is generic, i.e. it does not depend on the pair of Bell states that are chosen as entries of the convex combination.

\begin{figure}[H]
    \centering
    \includegraphics[scale =.3]{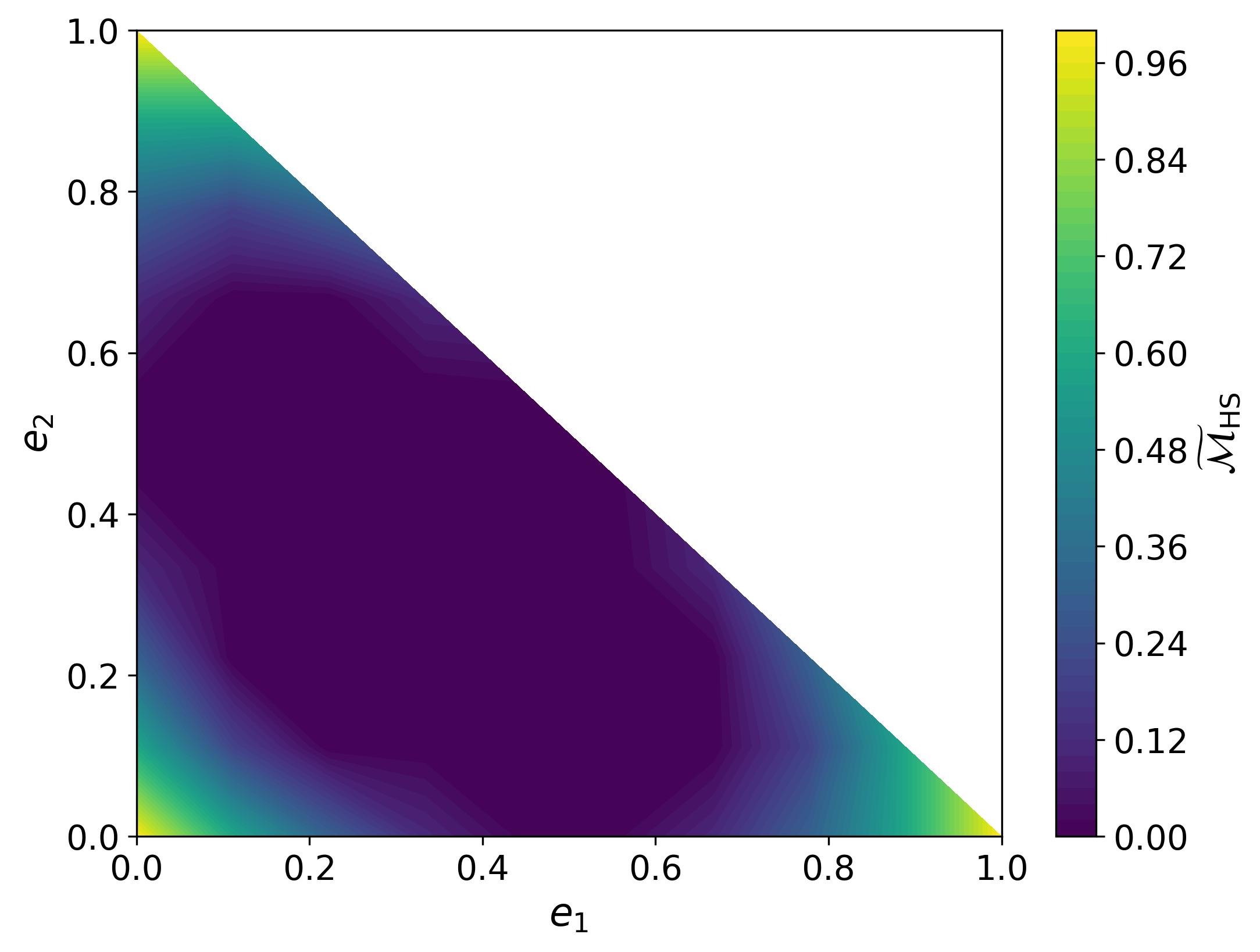}
    \caption{Normalized Hilbert-Schmidt measure $\widetilde{\mathcal{M}}_\mathrm{HS}$ of CHSH nonlocality for Bell-diagonal states as a function of the two free parameters $e_{1}$ and $e_{2}$, with $e_{4}=0$, $e_{3}=1-e_{1}-e_{2}$, and normalization coefficient $\mathcal{M}_\mathrm{HS}^{max}$. The maximum $\widetilde{\mathcal{M}}_\mathrm{HS}=1$ is reached in the three cases $(e_{1},e_{2},e_{3})=(1,0,0)$; $(e_{1},e_{2},e_{3})=(0,1,0)$; and $(e_{1},e_{2},e_{3})=(0,0,1)$. Each case corresponds to a different Bell state. The central region of minima is composed by local states, i.e. states with $\widetilde{\mathcal{M}}_\mathrm{HS}=0$. All plotted quantities are dimensionless.}
    \label{fig:HS2parameters}
\end{figure}

In the case of the Hilbert-Schmidt norm, Fig.~\ref{fig:HS2parameters} displays the corresponding nonlocality measure for a convex combination of two Bell states. 

\section{Higher dimensions and multipartite systems}
\label{sec:cglmp_inequality}
The CHSH scenario has been generalized to two-qudit systems, for which some relevant inequalities are the CGLMP inequalities \cite{Masanes2003,Collins2002} ($m=2, \Delta=d$), and to an arbitrary number of qubits, for $m=2$, $\Delta=2$ \cite{Mermin1990,Belinski1993,Zukowski2002} and for arbitrary $m$ \cite{Laskowski2004}. These inequalities are tight, and so they completely characterize the set of local correlations for the selected scenario.
For more complex scenarios, while it’s often the case that some Bell inequalities are known, they are generally not tight.

\subsection{Bipartite isotropic states}

An isotropic state has the form 
$\rho_{iso}=\omega\ket{\phi^+}\bra{\phi^+}+\frac{1-\omega}{d^2}\mathds{1}_d \otimes \mathds{1}_d$,  
where 
$\ket{\phi^+}\bra{\phi^+}=\sum_{ij=0}^{d-1}\ket{ii}\bra{jj}$,  
and 
$-\frac{1}{d^2-1}\leq \omega \leq 1$ \cite{Bertlmann2023}.  
In the case where we restrict to $0\leq \omega \leq 1$, this can be interpreted as a mixture of a maximally entangled state affected by white noise. If $\omega > \frac{1}{d+1}$, $\rho_{iso}$ is entangled.  

For any local isotropic state, the following result holds:

\noindent {\bf Proposition 3.} \textit{For any Bell inequality and for any jointly convex and unitarily invariant functional $D$ of pairs of two-qudit density matrices,} 
\begin{equation}
    \min_{\rho^{loc}\in \mathcal{L} }D(\rho_{iso},\rho^{loc})=\min_{\rho^{loc}_{iso}\in \mathcal{L} }D(\rho_{iso},\rho_{iso}^{loc}),
\end{equation}
\textit{where $\rho_{iso}$ is an isotropic state, $\rho^{loc}$ is a local state, and $\rho^{loc}_{iso}$ is a local isotropic state.}

\noindent
\textbf{Proof:} The proof is based on the following observations: 
\begin{enumerate}
    \item Obs.~1: Invariance of isotropic states under $U\otimes U^*$ transformations:
    \begin{equation}
        (U\otimes U^*) \, \rho_{iso} \, (U^\dagger \otimes {U^*}^\dagger) = \rho_{iso},
        \quad \forall \, U \ \mathrm{unitary}.
    \end{equation}

    \item Obs.~2: Local unitary transformations are resource non-generating (RNG) for nonlocality; in particular,
    \begin{equation}
        (U\otimes U^*) \, \rho^{loc} \, (U^\dagger \otimes {U^*}^\dagger) \in \mathcal{L},
    \end{equation}
    where $\mathcal{L}$ denotes the set of local states.

    \item Obs.~3: Any convex combination of local states is itself a local state.

    \item Any state $\rho$ can be transformed into an isotropic state by applying
    the twirling operation:
    \begin{equation}
        \int dU \, (U\otimes U^*) \, \rho \, (U^\dagger \otimes {U^*}^\dagger) = \rho_{iso}.
    \end{equation}
\end{enumerate}

From the unitary invariance of $D$, we have
\begin{eqnarray}
    D(\rho_{iso},\rho^{loc}_{iso}) 
    &=& D(V \, \rho_{iso} \, V^\dagger, V \, \rho^{loc} \, V^\dagger) \\
    &=& \int dV \, D(V \, \rho_{iso} \, V^\dagger, V \, \rho^{loc} \, V^\dagger) \\
    &\geq& D\Bigg(\int dV \, V \, \rho_{iso} \, V^\dagger, \int dV \, V \, \rho^{loc} \, V^\dagger \Bigg) \\
     &=& D(\rho_{iso}, \rho^{loc}_{iso})\,,
\end{eqnarray}
where, in going from the third line to the fourth, we defined $V=U\otimes U^*$.
Proposition 3 is general, since we do not specify any Bell inequality. In practice, for isotropic states, the set of local states that do not violate the CGLMP inequality is known \cite{Collins2002,Masanes2003}. This inequality is violated when $\omega>\frac{2}{I_d(QM)}$, with

\begin{equation}
    I_d(QM)=4d\sum_{k=0}^{[d/2]-1}\left(1-\frac{2k}{d-1}\right)(q_k-q_{-(k+1)})
\end{equation}
and
\begin{equation}
    q_k = 
    \frac{1}{2d^3\sin^2\left[\frac{\pi(k+\frac{1}{4})}{d}\right]}\,.
\end{equation}

When $m=2$ and $\Delta=d=2$, the only facet inequality reduces to CHSH. This inequality is violated when $\omega > \frac{1}{\sqrt{2}}$. For $m=2$ and $\Delta=d=3$, the facet inequalities correspond to the CGLMP inequality, which is violated when $\omega>(6\sqrt{3}-9)/2$.  

Here, we present the CGLMP nonlocality measures for isotropic states 
$\rho_{iso}=\omega \ket{\phi^+}\bra{\phi^+}+\frac{1-\omega}{d^2}\mathds{1}_d\otimes \mathds{1}_d$, 
for the different considered distances and assuming two measurements per party with $d$ outcomes:

\begin{equation}
    \mathcal{M}_\mathrm{HS}(\rho_{iso})=\sqrt{1-\frac{1}{d^2}}\left(\omega-\frac{2}{I_d(QM)}\right)
\end{equation}
\begin{equation}
    \mathcal{M}_\mathrm{Tr}(\rho_{iso})=\frac{2(d^2-1)}{d^2}\left(\omega-\frac{2}{I_d(QM)}\right)
\end{equation}

\begin{eqnarray}
\mathcal{M}_\mathrm{He} &=& 2 - \frac{2}{d}\Bigg[ (d^{2}-1)\sqrt{(1-\omega)\left(1-\frac{2}{I_d(QM)}\right)} \nonumber \\ 
&+& \qquad \sqrt{\big[(d^2-1)\omega + 1\big]\left[(d^2-1)\frac{2}{I_d(QM)} + 1\right]} \Bigg]
\end{eqnarray}

\begin{eqnarray}
    \mathcal{M}_{\mathrm{Re}} &=& \frac{(d^{2}-1)\omega + 1}{d^2} \log{\left( \frac{(d^2-1)\omega + 1}{d^2}\right)} + \frac{d^2-1}{d^2}\log{\left(\frac{1-\omega}{d^2}\right)} \nonumber \\
    &+& \frac{(d^{2}-1)\frac{2}{I_d(QM)} + 1}{d^2} \log{\left( \frac{(d^2-1)\frac{2}{I_d(QM)} + 1}{d^2}\right)} \nonumber \\
    &+& \frac{d^2-1}{d^2}\log{\left(\frac{1-\frac{2}{I_d(QM)}}{d^2}\right)}
\end{eqnarray}

\subsection{Bipartite Werner states}

The result obtained in the previous section can also be extended to Werner states in higher dimensions, i.e., states that are invariant under transformations of the form $U \otimes U$ \cite{Werner1989}:  

\noindent {\bf Proposition 4.} \textit{For any Bell inequality and for any jointly convex and unitarily invariant functional $D$, the closest local state to a Werner state is again a Werner state.} 

The proof follows exactly the same reasoning as for isotropic states. On the other hand, unlike the case of isotropic states, the precise value of the parameter that separates the nonlocal Werner states from the local ones is at present unknown. As a consequence, considering for instance states that are assured to be either CGLMP-local or CGLMP-nonlocal, our quantifiers provide, respectively, upper and lower bounds on the exact geometric measure of CGLMP nonlocality. In general, the full characterization of the set of local states for two-qudit systems in arbitrary Hilbert-space dimension requires a very ample and computationally complex effort, which will be the subject of a subsequent work.

\subsection{Multipartite systems}
Examples of Bell inequalities are also known in multipartite systems, such as the Mermin inequality~\cite{Mermin1990} and the Mermin-Ardehali-Belinskii-Klyshko (MABK) inequalities~\cite{Belinski1993}. In the case of $n$ qubits and two dichotomic measurements per party ($m=2$, $\Delta=2$), a complete characterization of all possible Bell inequalities has been obtained~\cite{Zukowski2002}.

Note that the results of the previous propositions also hold for generalized Werner states, defined as those states that are invariant under local unitaries $V \equiv \bigotimes_{i=1}^{n} U$, since the proof follows directly for the same reasons discussed above. Nevertheless, for $n>2$, Werner states are no longer simply given by a convex combination of their symmetric and antisymmetric parts~\cite{Eggeling2001}. Therefore, the study of Bell nonlocality in the multipartite case requires a separate and dedicated analysis which will be the subject of a forthcoming work.

\section{Discussion and outlook} \label{sec:conclusions}

We introduced a geometric measure for Bell nonlocality valid in any scenario, whose defining elements are the state vectors in Hilbert space and their distances from the set of local states. For any Bell scenario, we have identified the basic traits of the geometric framework, including the set of local states and the free operations \cite{Buscemi2012,Schmid2020b,Rosset2020,Lipka2021,Gallego2017}. 

The geometric approach is methodologically relevant because it enables a direct characterization and quantification of nonlocality in Hilbert space, thus placing it on the same conceptual footing as other fundamental measures of quantumness such as coherence, entanglement, steering, and quantum discord. Besides enabling the direct comparisons between nonlocality and different quantum resources in terms of inequalities and other hierarchical relations, the geometric framework allows in principle to investigate the problem of establishing whether sets of local operations exist that, by acting on maximally nonlocal states, allow to recover the entire Hilbert state space, in analogy with the case of coherence and entanglement. 

The quantification of Bell nonlocality for quantum states in Hilbert space is naturally introduced in terms of geometric measures, in close methodological analogy with the standard theory of geometric entanglement \cite{Wei2003,Uyanik2010} and the more recent geometric approaches to the quantification of other types of nonlocal correlations such as, e.g., the quantum discord \cite{Geomdiscord2016,Geomdiscord2017}. 

Geometric measures define a structure and an ordering in the space of quantum states, and we have provided a detailed investigation of such structure in general Bell scenarios for any number of parties, arbitrary finite Hilbert space dimension, and different classes of contractive distances (trace, Hellinger, Bures). In particular, we proved rigorously that for contractive distances the closest local state to a Werner (isotropic) state is still a Werner (isotropic) state, both in the bipartite and in the multipartite case. We have shown that an analogous result holds for two-qubit Bell-diagonal states. We have confronted the ordering provided by contractive geometric measures of nonlocality with the non-contractive Hilbert-Schmidt distance, and we have introduced the relative entropy of nonlocality as a further benchmark. 

In a hierarchy of increasing complexity, moving forward from the CHSH two-qubit scenario, we have determined the explicit expression of the geometric measures of nonlocality for isotropic two-qudit states relative to the CGLMP inequality in any Hilbert-space dimension, and we have outlined the general features of a geometric theory of multipartite nonlocality relative to the Mermin and the MABK inequalities.

In forthcoming works, for the bipartite case we plan to derive explicit analytical and numerical expressions of the geometric nonlocality of two-qudit Werner states in arbitrary dimension relative to the CGLMP inequality as well as in more general scenarios. For the multipartite case, we intend to compute the geometric measure of nonlocality of various classess of multipartite states, including generalized Werner states, in the Mermin, MABK, and more general scenarios \cite{Zukowski2002,Laskowski2004}.

Further generalizations and applications are conceivable along different directions, such as the identification of the set of local states and the ordering of nonlocal states in finite-dimensional Hilbert spaces of higher dimension, the extension of the geometric methods to different Bell inequalities both in the bipartite and in the multipartite case, the generalization to the field-theoretical setting of elementary particle physics, and the use of the Hilbert space-based measures of Bell nonlocality as tools in the investigation and characterization of quantum collective phenomena and quantum phase transitions. Indeed, the problem of characterizing the set of local states in arbitrary dimensions is known to be computationally hard, but by focusing on analytically tractable cases and leveraging known results on specific scenarios, it may be possible to provide concrete results and proofs that go beyond a conceptual proposal whenever partial knowledge of the local set is available or can be obtained by controlled approximation schemes.

Concerning the investigation of quantum matter, a characterization of the ground states of quantum many-body systems in terms of Hilbert-space-based measures of Bell nonlocality could provide novel and potentially profound insights into the degree of quantumness and quantumness hierarchies between different systems of quantum matter, in analogy with and complementing the very successful use of entanglement in condensed matter theory~\cite{Laflorencie2016,Illuminati2022,Illuminati2023}. Indeed, some recent preliminary studies on the structure of Bell nonlocality in quantum many-body systems hint at complex patterns of quantum nonlocality beyond entanglement \cite{Nonlocmanybody2015,Belldepthmanybody2019}, whose precise entropic and geometric quantification might provide important insights into the structure and hierarchy of quantum collective phenomena.

The Hilbert-space characterization and quantification that we have introduced pave the way for the study of several important questions concerning quantum nonlocality. In entanglement theory, it is known that different entanglement measures can give rise to different orderings of quantum states; for instance, the entropy-based entanglement of formation and the entanglement negativity based on the PPT criterion induce a different ordering on the set of entangled Gaussian states~\cite{Adesso2005}. Similar ordering issues might arise when determining the degree of Bell nonlocality of quantum states, as quantified by the different geometric or entropic measures that we have introduced in the present work. A further important question concerns the characterization and quantification of extremal nonlocality in mixed states. In entanglement theory, maximally entangled mixed states are typically defined at fixed energy, or fixed degrees of local and global entropies~\cite{Adesso2003,Adesso2004}. An investigation along similar lines could lead to the identification of the maximally nonlocal mixed states at fixed global and/or local purities. 

In our opinion, a particularly challenging open problem whose eventual solution would bear far-reaching consequences revolves around the question of whether the analogs of Werner states, i.e., mixed states that are local and yet entangled, exist in infinite-dimensional Hilbert spaces. In particular, it would be very interesting to determine the existence of Werner Gaussian states of two- and multimode continuous-variable systems, and to characterize their properties.

\section*{Acknowledgments} 
W.R. and M.T. are supported by JST Moonshot R\&D Grant No.~JPMJMS2061 and JST COI-NEXT Grant No.~JPMJPF2221. F.I. acknowledges funding from the Italian Ministry of University and Research, call PRIN PNRR 2022, project ``Harnessing topological phases for quantum technologies'', code P202253RLY, CUP D53D23016250001, and PNRR-NQSTI project ECoN: ``End-to-end long-distance entanglement in quantum networks", CUP J13C22000680006.

\appendix

\section{Two-qubit Bell-diagonal states and Werner states}
\label{appendix_1}

\subsection{Bell-diagonal states}

Given the Bell basis formed by the four maximally entangled Bell states 
\begin{equation*}
\ket{\Psi_{1}} = \frac{\ket{00}+\ket{11}}{\sqrt{2}} \, , \qquad \ket{\Psi_{2}} = \frac{\ket{00}-\ket{11}}{\sqrt{2}} \, ,
\end{equation*}
\begin{equation}
\ket{\Psi_{3}} = \frac{\ket{01}+\ket{10}}{\sqrt{2}} \, , \qquad \ket{\Psi_{4}} = \frac{\ket{01}-\ket{10}}{\sqrt{2}} \, ,
\label{Bellstates}
\end{equation}
their convex combinations define the Bell-diagonal states:
\begin{equation}
\rho_{\mathrm{BD}} = \displaystyle\sum_{i=1}^{4}e_{i} \ket{\Psi_{i}} 
\bra{\Psi_{i}} \, ,
\label{generic_BD}
\end{equation}
where $\{ 0 \leq e_i \leq 1 \}$, with $\displaystyle\sum_{i=1}^{4}e_{i} = 1$. By resorting to the notation of Eq.~\eqref{representation}, a generic Bell-diagonal state  Eq.~\eqref{generic_BD} can be rewritten as
\begin{equation}
\rho_{\mathrm{BD}} = \frac{1}{4} \bigg[ \mathds{1}_{2} \otimes \mathds{1}_{2} + \displaystyle\sum_{i=1}^{3} a_{i} \sigma_{i} \otimes \sigma_{i}\bigg] \, ,
\end{equation}
where $\mathds{1}_{2}$ denotes the identity matrix in two dimensions. Matrix $\alpha$ of Eq.~\eqref{alpha_matrix_generic} can then be recast in the form
\begin{equation}
    \alpha_{\mathrm{BD}} = 
    \left[
    \begin{array}{cccc}
    1 & 0 & 0 & 0\\
    0 & a_1 & 0 & 0\\
    0 & 0 & a_2 & 0\\
    0 & 0 & 0 & a_3
    \end{array}
    \right] \,.
    \label{BD_matrix}
\end{equation}

\begin{figure}
    \centering
    \includegraphics[scale = .35]{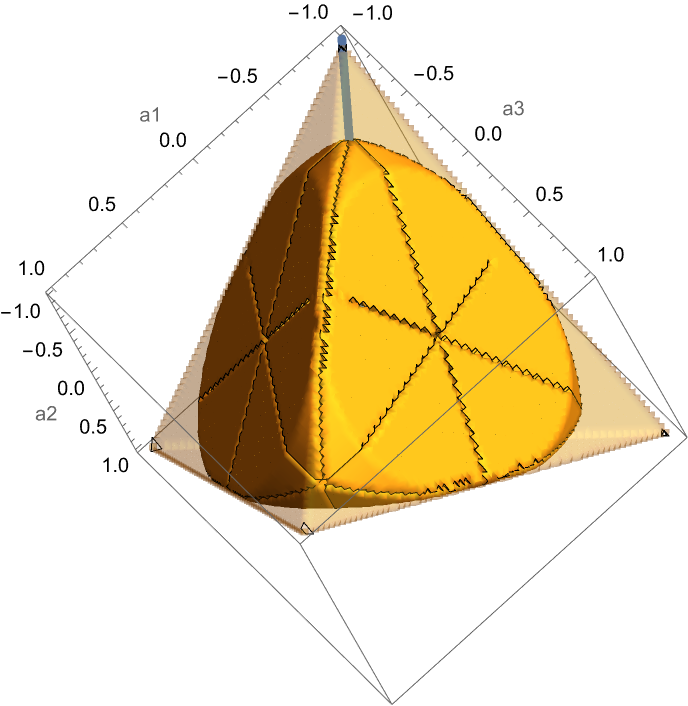}
    \caption{Semitransparent tetrahedron: set of Bell-diagonal states. The set is parameterized by the matrix elements $(a_1, a_2, a_3)$ in Eq.~\eqref{BD_matrix}. Solid figure inside the tetrahedron: set of local states. Solid grey line: set of Werner states.}
    \label{fig:pic}
\end{figure}

Collecting the above results, for any Bell-diagonal state it is rather straightforward to derive the following formula relating the coefficients $\vec{a}=(1,a_1,a_2,a_3)^T$ of the matrix $\alpha$ and the eigenvalues $\vec{e}=(e_1,...,e_4)^T$ of the density matrix:
\begin{equation}
\left[
\begin{array}{c}
1\\
a_1\\
a_2\\
a_3
\end{array}
\right]
=
\left[
\begin{array}{cccc}
1 & 1 & 1 & 1\\
1 & 1 & -1 & -1\\
1 & -1 & 1 & -1\\
-1 & 1 & 1 & -1
\end{array}
\right]
\left[
\begin{array}{c}
e_1\\
e_2\\
e_3\\
e_4
\end{array}
\right] \,.
\label{matrix}
\end{equation}

The matrix in the above expression is unitary upon multiplication by $1/2$. It follows that

\begin{equation}
\left[
\begin{array}{c}
e_1\\
e_2\\
e_3\\
e_4
\end{array}
\right]
=
\frac{1}{4}
\left[
\begin{array}{cccc}
1 & 1 & 1 & -1\\
1 & 1 & -1 & 1\\
1 & -1 & 1 & 1\\
1 & -1 & -1 & -1
\end{array}
\right]
\left[
\begin{array}{c}
1\\
a_1\\
a_2\\
a_3
\end{array}
\right] \,.
\label{eigenvalues}
\end{equation}

so that the positivity and normalization of the sum of the eigenvalues allow us to represent the set of all Bell-diagonal states parameterized by the three-dimensional vector $(a_1,a_2,a_3)$ as shown in Fig. (\ref{fig:pic}). 

\subsection{Werner states}

A particular subclass of the Bell-diagonal states are the Werner states $\rho_w$, described by one free parameter $|a_1|=|a_2|=|a_3| \equiv w$. Formally, a Werner state is a convex combination of the maximally mixed state with any one of the four Bell states. This class of states, first introduced by R. F. Werner \cite{Werner1989}, is important as it provides the paramount example of a set of states that for $\frac{1}{3} \leq w \leq \frac{1}{\sqrt{2}}$ are entangled but nevertheless local.

\section{Bell-diangonal states}
\label{appendix_2}
\subsection{Hilbert-Schmidt distance}

Combining the above with Proposition 2, Eq.~\eqref{Proposition2}, the minimum of the squared HS distance is determined as 
 \begin{eqnarray}
     \min_{\rho_{\mathrm{BD}}^{loc} \in \mathcal{L}} \left\| \rho_{\mathrm{BD}}-\rho_{\mathrm{BD}}^{loc} \right\|_{2} = \min_{a_{i}^{loc}}\frac{1}{2} \sqrt{ \displaystyle\sum_{i=1}^{3} (a_{i}-a_{i}^{loc})^{2}}  \, ,
     \label{HS_BD}
 \end{eqnarray}
 with the provision that the states $\rho_{\mathrm{BD}}$ belong to the surface Eq.~\eqref{Surface_HS}. 

Considering first the case $|a_{1}^{loc}|,|a_{2}^{loc}|>|a_{3}^{loc}|$, we can proceed to determine the minimum
by introducing a Lagrange multiplier $\lambda$. The associated Lagrange variational (squared) function reads
\begin{eqnarray}
 \label{Lagrange_HS}
     \Gamma(a_{1}^{loc},a_{2}^{loc},a_{3}^{loc},\lambda) \equiv \displaystyle\sum_{i=1}^{3} (a_{i}-a_{i}^{loc})^{2} + \lambda[(a_{1}^{loc})^{2}+(a_{2}^{loc})^{2}-1] \, .
 \end{eqnarray}
The vanishing of the gradient of $\Gamma$ leads to
\begin{equation}
\left\{
\begin{array}{rcl}
(a_{1} - a_{1}^{loc}) + \lambda a_{1}^{loc} &=& 0, \\
(a_{2} - a_{2}^{loc}) + \lambda a_{2}^{loc} &=& 0, \\
(a_{3} - a_{3}^{loc}) &=& 0, \\
(a_{1}^{loc})^{2} + (a_{2}^{loc})^{2} - 1 &=& 0.
\end{array}
\right.
\end{equation}

Denoting the minima by $\tilde{a}_{i}^{loc}$, the solution reads
\begin{equation*}
a_{1}^{loc} = \pm \frac{a_{1}} {\sqrt{a_{1}^{2}+a_{2}^{2}}},
\end{equation*}
\begin{equation*}
a_{2}^{loc} = \pm \frac{a_{2} }{\sqrt{a_{2}^{1}+a_{2}^{2}}} ,
\end{equation*}
\begin{equation}
a_{3}^{loc} = a_{3} .
\label{lagrangehspoints}
\end{equation}
By Eq.~\eqref{HS_BD} the (un-normalized) HS geometric measure of CHSH nonlocality for Bell-diagonal states is thus
\begin{equation}
\mathcal{M}_\mathrm{HS}(\rho_{\mathrm{BD}}) = \frac{1}{2}\sqrt{\sum\limits_{i=x,y,z}(a_i-a_i^{loc})^2}=\frac{1}{2}\left(\sqrt{a_{1}^2+a_{2}^2}\mp1\right)
\label{lagrangehs}
\end{equation}
whose minimum clearly corresponds to the minus sign (the plus sign in (\ref{lagrangehspoints})), since $a_x^2+a_y^2\geq1$. It should be noted that the solution (\ref{lagrangehspoints}) is acceptable only if it satisfies the condition $|a_1^{loc}|,|a_2^{loc}|>|a_3^{loc}|$.
The remaining cases $|a_{2}^{loc}|,|a_{3}^{loc}| > |a_{1}^{loc}|$ and $|a_{1}^{loc}|,|a_{3}^{loc}|> |a_{2}^{loc}|$ can be immediately found by symmetry, and the global minimum is determined over all possible cases, provided that the corresponding $|a_i^{loc}|$ are correctly ordered. 

It may happen that none of the three cases gives rise to a coherent solution; in this case, it is not possible to express the HS measure in the form (\ref{lagrangehs}). This is the case of Werner states, where $|a_i|=w$, but the solution (\ref{lagrangehspoints}) implies $|a_1^{loc}|,|a_2^{loc}|=1/\sqrt{2}<|a_3^{loc}|$. In this case the correct solution is indeed different from (\ref{lagrangehs}) (see (\ref{measurehswerner})). 

 Using Eqs. \eqref{matrix} and \eqref{eigenvalues} the transformation from the matrix elements $\{ a_i \}$ to the eigenvalues $\{e_i\}$ of the density matrix follows immediately. The behavior of the normalized HS measure of CHSH nonlocality $\widetilde{\mathcal{M}}_\mathrm{HS} = \mathcal{M}_\mathrm{HS}/\mathcal{M}_\mathrm{HS}^{max}$ for Bell-diagonal states as a function of the eigenvalues $e_1$, $e_2$ in the case of $e_{4}=0$ is reported in Fig. \ref{fig:HS2parameters}.

\subsection{Hellinger distance}
Each of the eigenvalues $\{e_{i}^{k}\}$ of a density matrix $\rho_k$ is the probability that a system in state $\rho_{k}$ collapses in the eigenstate relative to that eigenvalue. For a Bell-diagonal state $\rho_{\mathrm{BD}}$ the eigenstates are obviously the four Bell states. Recalling Proposition 2, Eq.~\eqref{Proposition2}, and Eq.~\eqref{HellingerDistance} the minimization process on the Hellinger distance squared yields the Hellinger geometric measure of CHSH nonlocality $\mathcal{M}_\mathrm{He}(\rho_{\mathrm{BD}})$ for Bell-diagonal states:
\begin{equation}
   \mathcal{M}_\mathrm{He}(\rho_{\mathrm{BD}}) = \min_{\rho_{\mathrm{BD}}^{loc} \in \mathcal{L}}D^2_\mathrm{He}(\rho_{\mathrm{BD}}, \rho_{\mathrm{BD}}^{loc}) =  \min_{\rho^{loc} \in \mathcal{L}} \left\{ \displaystyle\sum_{i=1}^{4} \bigg( \sqrt{e_{i}}-\sqrt{e_{i}^{loc}} \bigg)^{2} \right\} \, .
   \label{MinimumHellinger}
\end{equation}
Using Eq.~\eqref{eigenvalues} we can rewrite the condition Eq.~\eqref{Surface_HS} in terms of the eigenvalues 
$\{e_{i}^{loc}\}$. Fixing, for instance, the case $|a_{1}|,|a_{2}|>|a_{3}|$, one has:
\begin{equation}
\label{ipervolume_nonlocal}
(e_{1} - e_{3})^{2} + (e_{2}-e_{4})^{2} \, > \, \frac{1}{2} \, .
\end{equation}
For probabilities, their positivity $e_{i}>0$ and normalization condition $\displaystyle\sum_{i}e_{i}=1$ 
define a clove of a unitary $4-$sphere in the $\sqrt{e_{i}}-$space. Therefore, the set of nonlocal states with $|a_{1}|,|a_{2}| > |a_{3}|$ is represented in the $\sqrt{e_{i}}-$space by the intersection of the hypervolume Eq.~\eqref{ipervolume_nonlocal} with the unitary semi-sphere centered in the origin. Since the distance in this space is the Euclidean one, the minimum in Eq.~\eqref{MinimumHellinger} in the case $|a_{1}^{loc}|,|a_{2}^{loc}| > |a_{3}^{loc}|$ is determined by imposing the constraint 
\begin{equation}
(e_{1}^{loc} - e_{3}^{loc})^{2} + (e_{2}^{loc}-e_{4}^{loc})^{2} = \frac{1}{2} \, .
\label{constraint_probability}
\end{equation}
Recalling Eq.~\eqref{HellingerDistance}, the above relation identifying the region of separation between local and nonlocal states is obtained when $\sum_{i=1}^{4}\sqrt{e_{i}e_{i}^{loc}}$ is maximum. We can thus proceed by the method of Lagrange multipliers and introduce the following Lagrangian function
\begin{eqnarray}
\Gamma (e_{i}^{loc}, \lambda_{1}, \lambda_{2}) &\equiv& \displaystyle\sum_{i=1}^{4}\sqrt{e_{i}e_{i}^{loc}} + \lambda_{1}\bigg[(e_{1}^{loc} - e_{3}^{loc})^{2} + (e_{2}^{loc}-e_{4}^{loc})^{2}-\frac{1}{2}\bigg] \nonumber \\
&+& \lambda_{2}\left( \displaystyle\sum_{i=1}^{4}e_{i}^{loc}-1 \right) \, .
\end{eqnarray}
Taking the gradient of $\Gamma$ and imposing its vanishing yields
\begin{equation}
\left\{
\begin{array}{l}
\frac{1}{2}\sqrt{\frac{e_{1}}{e_{1}^{loc}}} + 2 \lambda_{1}(e_{1}^{loc} - e_{3}^{loc}) + \lambda_{2} = 0, \\[6pt]
\frac{1}{2}\sqrt{\frac{e_{2}}{e_{2}^{loc}}} + 2 \lambda_{1}(e_{2}^{loc} - e_{4}^{loc}) + \lambda_{2} = 0, \\[6pt]
\frac{1}{2}\sqrt{\frac{e_{3}}{e_{3}^{loc}}} - 2 \lambda_{1}(e_{1}^{loc} - e_{3}^{loc}) + \lambda_{2} = 0, \\[6pt]
\frac{1}{2}\sqrt{\frac{e_{4}}{e_{4}^{loc}}} - 2 \lambda_{1}(e_{2}^{loc} - e_{4}^{loc}) + \lambda_{2} = 0, \\[6pt]
e_{1}^{loc} + e_{2}^{loc} + e_{3}^{loc} + e_{4}^{loc} = 1, \\[6pt]
(e_{1}^{loc} - e_{3}^{loc})^{2} + (e_{2}^{loc} - e_{4}^{loc})^{2} = \frac{1}{2}.
\end{array}
\right.
\end{equation}

From the above relations one immediately determines the Lagrange multipliers:
\begin{equation}
\lambda_{1} = -\frac{\sqrt{\frac{e_{1}}{e_{1}^{loc}}}-\sqrt{\frac{e_{3}}{e_{3}^{loc}}}}{8(e_{1}^{loc}-e_{3}^{loc})} = -\frac{\sqrt{\frac{e_{2}}{e_{2}^{loc}}}-\sqrt{\frac{e_{4}}{e_{4}^{loc}}}}{8(e_{2}^{loc}-e_{4}^{loc})} \, ,
\end{equation}
\begin{equation}
\lambda_{2} = -\frac{1}{4}\bigg( \sqrt{\frac{e_{1}}{e_{1}^{loc}}}+\sqrt{\frac{e_{3}}{e_{3}^{loc}}}  \bigg) = -\frac{1}{4}\bigg(\sqrt{\frac{e_{2}}{e_{2}^{loc}}}+\sqrt{\frac{e_{4}}{e_{4}^{loc}}}   \bigg) \, ,
\end{equation}
so that
\begin{equation}
\left\{
\begin{array}{l}
\sqrt{\frac{e_{1}}{e_{1}^{loc}}} + \sqrt{\frac{e_{3}}{e_{3}^{loc}}} = \sqrt{\frac{e_{2}}{e_{2}^{loc}}} + \sqrt{\frac{e_{4}}{e_{4}^{loc}}} \, , \\[6pt]
\bigg( \sqrt{\frac{e_{1}}{e_{1}^{loc}}} - \sqrt{\frac{e_{3}}{e_{3}^{loc}}} \bigg)(e_{2}^{loc} - e_{4}^{loc}) = \bigg( \sqrt{\frac{e_{2}}{e_{2}^{loc}}} - \sqrt{\frac{e_{4}}{e_{4}^{loc}}} \bigg)(e_{1}^{loc} - e_{3}^{loc}) \, , \\[6pt]
e_{1}^{loc} + e_{2}^{loc} + e_{3}^{loc} + e_{4}^{loc} = 1 \, , \\[6pt]
(e_{1}^{loc} - e_{3}^{loc})^{2} + (e_{2}^{loc} - e_{4}^{loc})^{2} = \frac{1}{2} \, .
\end{array}
\right.
\end{equation}

The remaining cases can be treated symmetrically; the absolute minimum is determined by the smallest value with respect to all possible cases. In the case of Bell states, e.g. $e_{1}=1$, $e_{2}=e_{3}=e_{4}=0$, it is straightforward to verify that the minimum is reached for $e_{1}^{loc} = \frac{1+3t}{4}$, $e_{2}^{loc} = e_{3}^{loc} = e_{4}^{loc} = \frac{1-t}{4}$ , with $t=\frac{1}{\sqrt{2}}$. Therefore, as it must be, we find that the maximum value $\mathcal{M}_{\mathrm{He}}^{max}$ of the Hellinger geometric measure of CHSH nonlocality Eq.~\eqref{MinimumHellinger} coincides with that of the Hellinger measure for Werner states, Eq.~\eqref{Werner_Hellinger}, with the Werner parameter $w=1$.

On the other hand, we see that the above minimization procedure cannot always yield a closed formula of the Hellinger measure of CHSH nonlocality for generic Bell-diagonal states. This is implied by the type of bound on the set of local states and the form of the function that one needs to minimize. Indeed, the above algebraic systems yield that in order to find the minimum one must determine the roots of a polynomial of the fifth order. Therefore, from the Abel-Ruffini-Seralian theorem, it follows that it is impossible to obtain a closed analytical formula by quadratures for the roots in terms of the coefficients of the polynomial. Of course, the solutions can be determined to any desired degree of precision by standard approximation procedures such as, e.g., the Newton-Raphson method or the Laguerre iterative scheme. Alternatively, we can proceed analytically in special but relevant instances, such as the Bell states or the Bell-diagonal states with one free parameter.

\subsection{Bures distance}
In analogy with the case of Werner states, in the case of Bell-diagonal states as well, the density operators commute, and the Bures and Hellinger distances coincide, so that
\begin{equation}
 \mathcal{M}_{\mathrm{Bu}}(\rho_{\mathrm{BD}}) = \mathcal{M}_{\mathrm{He}}(\rho_{\mathrm{BD}}) \, .    
\end{equation}

\subsection{Relative entropy}

In the case of the relative entropy, we assume that in order to determine the minimum in Eq.~\eqref{relativeEntropy} we can continue to impose the constraint Eq.~\eqref{constraint_probability}. In other words, we expect that for the relative entropy of nonlocality of Bell-diagonal states, the closest local state belongs to the same surface as for the Hellinger and Bures measures. Assuming the above hypothesis, the minimum in Eq.~\eqref{relativeEntropy} is realized when 
\begin{equation}
\displaystyle\sum_{i=1}^{3} e_{i} \log_{2}{e_{i}^{loc}}
\end{equation}
achieves its maximum.\\
Fixing, as usual, to begin with the case $|a_{1}|,|a_{2}|>|a_{3}|$, the Lagrange function associated with the Lagrange multipliers takes the form
\begin{eqnarray}
\Gamma(e_{i}^{loc}, \lambda_{1}, \lambda_{2}) &\equiv& \displaystyle\sum_{i=1}^{4} e_{i} \log_{2}{e_{i}^{loc}}+ \lambda_{1}[(e_{1}^{loc} - e_{3}^{loc})^{2} + (e_{2}^{loc}-e_{4}^{loc})^{2}-\frac{1}{2}] \nonumber \\
&+& \lambda_{2}\left( \displaystyle\sum_{i=1}^{4}e_{ij}^{loc}-1 \right) \, ,
\end{eqnarray}
which leads to
\[
\left\{
\begin{array}{l}
\displaystyle
\frac{e_{1}}{e_{1}^{loc}\ln 2} + 2 \lambda_{1} (e_{1}^{loc} - e_{3}^{loc}) + \lambda_{2} = 0 , \\[6pt]

\displaystyle
\frac{p_{01}}{e_{2}^{loc}\ln 2} + 2 \lambda_{1} (e_{2}^{loc} - e_{4}^{loc}) + \lambda_{2} = 0 , \\[6pt]

\displaystyle
\frac{e_{3}}{e_{3}^{loc}\ln 2} - 2 \lambda_{1} (e_{1}^{loc} - e_{3}^{loc}) + \lambda_{2} = 0 , \\[6pt]

\displaystyle
\frac{e_{4}}{e_{4}^{loc}\ln 2} - 2 \lambda_{1} (e_{2}^{loc} - e_{4}^{loc}) + \lambda_{2} = 0 , \\[6pt]

e_{1}^{loc} + e_{2}^{loc} + e_{3}^{loc} + e_{4}^{loc} = 1 , \\[6pt]

(e_{1}^{loc} - e_{2}^{loc})^{2} + (e_{3}^{loc} - e_{4}^{loc})^{2} = \frac{1}{2} .
\end{array}
\right.
\]

It is straightforward to show that the above relations reduce to
\[
\left\{
\begin{array}{l}
\displaystyle
\frac{e_{1}}{e_{1}^{loc}} + \frac{e_{3}}{e_{3}^{loc}} = \frac{e_{2}}{e_{2}^{loc}} + \frac{e_{4}}{e_{4}^{loc}} , \\[6pt]
\displaystyle
\bigg( \frac{e_{3}}{e_{3}^{loc}} - \frac{e_{1}}{e_{1}^{loc}} \bigg)(e_{2}^{loc} - e_{4}^{loc}) = \bigg( \frac{e_{4}}{e_{4}^{loc}} - \frac{e_{2}}{e_{2}^{loc}} \bigg)(e_{1}^{loc} - e_{3}^{loc}) , \\[6pt]
e_{1}^{loc} + e_{2}^{loc} + e_{3}^{loc} + e_{4}^{loc} = 1 , \\[6pt]
(e_{1}^{loc} - e_{2}^{loc})^{2} + (e_{3}^{loc} - e_{4}^{loc})^{2} = \frac{1}{2} .
\end{array}
\right.
\]

In analogy with the case of the Hellinger measure, the above algebraic system is of a higher order and not amenable to a solution by elementary quadratures. Therefore, in the following subsection, we will consider the particular instance of one-parameter Bell-diagonal states that allow for the full analytical evaluation of the entire set of geometric and entropic measures of CHSH nonlocality.

\section{Numerical Methods for the Estimation of the CHSH-Nonlocality Measures}
\label{app:numerical}

\subsection{Geometric measures}

In this section, we describe the procedures used to estimate the geometric and entropic nonlocality measures for Bell-diagonal states in cases where closed analytical formulas are not available. Recall that Bell-diagonal states are convex combinations of the four Bell states and are therefore characterized by three free parameters, namely the independent probabilities $e_{1}$, $e_{2}$, and $e_{3}$, since the fourth probability is fixed by normalization as $e_{4} = 1 - e_{1} - e_{2} - e_{3}$. Alternatively, one can adopt the parametrization \eqref{Bell-diagonal states in termini delle sigma} in terms of the vector $\mathbf{a} = (a_{1}, a_{2}, a_{3})$. For the numerical analysis, the choice of parametrization can be made according to convenience, since for different distances a suitable parametrization can simplify the minimization procedure.

\paragraph{Symmetric families.}
For Werner and isotropic states, we prove that the closest local state remains within the same family. This property allows us to derive explicit analytical expressions for the Hilbert-Schmidt, Hellinger, Bures, trace, and relative entropy measures of CHSH nonlocality (see Sections~\ref{sec:geometric_measures} and~\ref{sec:cglmp_inequality} of the main text). The formulas obtained for the two-qubit case are also employed as benchmarks to validate the numerical procedures described below.

\paragraph{Bell-diagonal states.}
For general Bell-diagonal states, the minimization cannot be solved analytically, since it is not guaranteed a priori that the candidate minimum $\rho_{\mathrm{loc}}$ is a physical density operator. In particular, its positivity must be explicitly verified. The local set $\sigma$ is convex, so in principle the problem is convex once expressed in a suitable parametrization. In practice, however, the locality constraints lead to nonlinear and piecewise boundary conditions, which make direct global minimization unreliable. 

To provide concrete estimation methods, we implemented a numerical strategy based on the \texttt{SciPy} optimization library. Specifically: 
\begin{itemize}
  \item The parameters of Bell-diagonal states are expressed either in terms of the probabilities $e_i$ or the correlators $(a_1,a_2,a_3)$, with explicit routines to convert between the two parameterizations.
  \item Locality constraints are encoded through inequalities on the $t_i$ or $e_i$, ensuring that the candidate $\rho^{loc} \in \mathcal{L}$ corresponds to a valid density operator. 
  \item Since the boundary of $\sigma$ consists of disjoint surfaces, the minimization is performed separately on each surface, and the smallest value is retained. This piecewise approach improves robustness. 
  \item Optimization is carried out with \texttt{minimize} (SLSQP or \texttt{trust-constr}), with bounds $[-1,1]$ on correlators and $[0,1]$ on probabilities. Multiple initial seeds (symmetric choices and random perturbations) are used to avoid local traps. 
  \item For entropic measures, the minimization is performed directly in the probability space, with constraints enforcing normalization and positivity. 
\end{itemize}

\subsection{Numerical procedure}

For Bell-diagonal states, we considered two different restricted families.  
First, we studied convex combinations of two Bell states, which depend on a single free parameter.  
Second, we analyzed convex combinations of three Bell states, which involve two free parameters, with the fourth probability fixed at zero by construction.  
In the numerical minimization, the Hilbert-Schmidt measure is evaluated in the $(a_1,a_2,a_3)$ parametrization, while for the other measures (Hellinger, Bures, trace, and relative entropy) the probability parametrization $(e_i)$ is used.  
Since Werner states form a subclass of Bell-diagonal states in the two-qubit case, we use the available analytical formulas for this family as benchmarks to validate our numerical procedures.  
This strategy provides concrete and stable estimates of the nonlocality measures and enables the generation of contour plots over the corresponding parameter spaces.

\subsection{Validation and accuracy}

To ensure correctness:
\begin{enumerate}
  \item We benchmark against the analytical solutions for Werner and isotropic states, reproducing the exact values within tolerance $<10^{-6}$. 
  \item We check grid convergence by refining the discretization (e.g.\ $n=10,20,50$ points per parameter) and verifying stability of the estimates. 
  \item We verify that different initializations lead to the same minimum, confirming consistency of the optimization. 
\end{enumerate}

All Python codes used for these computations are available on GitHub: 
\href{https://github.com/2zanfardino/geometric_nonlocality}{https://github.com/2zanfardino/geometric\_nonlocality}. 
Typical execution times for two-parameter contour plots (grid size $50 \times 50$) are a few minutes on a standard workstation.

\end{document}